\documentclass[twocolumn,floatfix,times,trackchanges,twocolappendix]{aastex63}

\def\sun{\hbox{$\odot$}}
\def\earth{\hbox{$\oplus$}}
\def\lesssim{\mathrel{\hbox{\rlap{\hbox{%
 \lower4pt\hbox{$\sim$}}}\hbox{$<$}}}}
\def\gtrsim{\mathrel{\hbox{\rlap{\hbox{%
 \lower4pt\hbox{$\sim$}}}\hbox{$>$}}}}

\def\arcs{\hbox{$^{\prime\prime}$}}

\def\farcs{\hbox{$.\!\!^{\prime\prime}$}}

\newcommand{\mJyperbeam}{\mbox{~mJy~beam$^{-1}$}}

\newcommand{\kmpers}{\mbox{~km~s$^{-1}$}}

\newcommand{\voff}{$v_\mathrm{off}$}
\newcommand{\frest}{$f_\mathrm{rest}$}
\newcommand{\Aij}{$A_\mathrm{ij}$}
\newcommand{\Eu}{$E_\mathrm{u}$}
\newcommand{\gu}{$g_\mathrm{u}$}
\newcommand{\vwidth}{$\delta v$}

\newcommand{\Trot}{$T_\mathrm{rot}$}
\newcommand{\Ntot}{$N_\mathrm{tot}$}
\newcommand{\Lsun}{$L_\odot$}
\newcommand{\Msun}{$M_\odot$}

\newcommand{\Mstar}{$M_\mathrm{\star}$}

\newcommand{\Tex}{$T_\mathrm{ex}$}

\usepackage{CJKutf8}
\usepackage{graphicx}
\usepackage{longtable} \LTcapwidth=\textwidth
\usepackage{multirow}
\usepackage{subfigure}
\usepackage{ulem}
\usepackage{soul}
\usepackage{lipsum, babel}
\usepackage{enumitem}
\usepackage{amsmath}
\usepackage{euscript}
\usepackage{verbatim}
\usepackage{hyperref}

\setstcolor{red}

\submitjournal{ApJ}

\shorttitle{HOPS-288, a COMs Lab.}
\shortauthors{Hsu et al.}
\graphicspath{{./}{figures/}}

\begin{document}

\begin{CJK*}{UTF8}{bsmi}
\title{HOPS-288: A Laboratory for Complex Organics in Proto-binary/multiple Systems}

\author[0000-0002-1369-1563]{Shih-Ying Hsu}
\email{seansyhsu@gmail.com}
\affiliation{Institute of Astronomy and Astrophysics, Academia Sinica, No.1, Sec. 4, Roosevelt Rd, Taipei 106319, Taiwan (R.O.C.)}

\author[0009-0001-6486-6909]{Nadia M. Murillo}
\affiliation{Instituto de Astronom\'ia, Universidad Nacional Aut\'onoma de M\'exico, AP106, Ensenada CP 22830, B. C., M\'exico}

\author[0000-0002-3024-5864]{Chin-Fei Lee}
\affiliation{Institute of Astronomy and Astrophysics, Academia Sinica, No.1, Sec. 4, Roosevelt Rd, Taipei 106319, Taiwan (R.O.C.)}

\author[0000-0002-6773-459X]{Doug Johnstone}
\affiliation{NRC Herzberg Astronomy and Astrophysics, 5071 West Saanich Rd, Victoria, BC, V9E 2E7, Canada}
\affiliation{Department of Physics and Astronomy, University of Victoria, Victoria, BC, V8P 5C2, Canada}

\author[0000-0002-5507-5697]{Tien-Hao Hsieh}
\affil{Taiwan Astronomical Research Alliance (TARA), Taiwan (R.O.C.)}
\affiliation{Institute of Astronomy and Astrophysics, Academia Sinica, No.1, Sec. 4, Roosevelt Rd, Taipei 106319, Taiwan (R.O.C.)}

\author[0000-0001-9304-7884]{Naomi Hirano}
\affiliation{Institute of Astronomy and Astrophysics, Academia Sinica, No.1, Sec. 4, Roosevelt Rd, Taipei 106319, Taiwan (R.O.C.)}

\author[0000-0002-9574-8454]{Leonardo Bronfman}
\affiliation{Departamento de Astronom\'{i}a, Universidad de Chile, Casilla 36-D, Santiago, Chile}

\author[0000-0001-5653-3584]{Yo-Ling Chuang}
\altaffiliation{Center of Astronomy and Gravitation, National Taiwan Normal University, 88 Sec.4 Ting-Chou Rd., Taipei 116, Taiwan, ROC}
\affiliation{Institute of Astronomy and Astrophysics, Academia Sinica, No.1, Sec. 4, Roosevelt Rd, Taipei 106319, Taiwan (R.O.C.)}

\author[0000-0002-5881-3229]{David J. Eden}
\affiliation{Department of Physics, University of Bath, Claverton Down, Bath BA2 7AY, UK}
\affiliation{Armagh Observatory and Planetarium, College Hill, Armagh BT61 9DB, UK}

\author[0000-0003-1275-5251]{Shanghuo Li}
\affiliation{School of Astronomy and Space Science, Nanjing University, 163 Xianlin Avenue, Nanjing 210023, People’s Republic of China}
\affiliation{Key Laboratory of Modern Astronomy and Astrophysics (Nanjing University), Ministry of Education, Nanjing 210023, People's Republic of China}

\author[0000-0002-6868-4483]{Sheng-Jun Lin}
\affiliation{Institute of Astronomy and Astrophysics, Academia Sinica, No.1, Sec. 4, Roosevelt Rd, Taipei 106319, Taiwan (R.O.C.)}

\author[0000-0002-6529-202X]{Mark G. Rawlings}
\affiliation{Gemini Observatory/NSF NOIRLab, 670 N. A'ohoku Place, Hilo, HI 96720, USA}

\author[0000-0002-8149-8546]{Ken'ichi Tatematsu}
\affiliation{Nobeyama Radio Observatory, National Astronomical Observatory of Japan, National Institutes of Natural Sciences, 462-2 Nobeyama, Minamimaki, Minamisaku, Nagano 384-1305, Japan}
\affiliation{Astronomical Science Program, The Graduate University for Advanced Studies, SOKENDAI, 2-21-1 Osawa, Mitaka, Tokyo 181-8588, Japan}

\author[0000-0012-3245-1234]{Sheng-Yuan Liu}
\affiliation{Institute of Astronomy and Astrophysics, Academia Sinica, No.1, Sec. 4, Roosevelt Rd, Taipei 106319, Taiwan (R.O.C.)}

\author[0000-0002-9774-1846]{Huei-Ru Vivien Chen}
\affiliation{Department of Physics and Institute of Astronomy, National Tsing Hua University, Hsinchu, 30013, Taiwan}

\author[0000-0003-2412-7092]{Kee-Tae Kim}
\affiliation{Korea Astronomy and Space Science Institute (KASI), 776 Daedeokdae-ro, Yuseong-gu, Daejeon 34055, Republic of Korea}

\author[0000-0002-4336-0730]{Yi-Jehng Kuan}
\affiliation{Department of Earth Sciences, National Taiwan Normal University, Taipei, Taiwan (R.O.C.)}
\affiliation{Institute of Astronomy and Astrophysics, Academia Sinica, No.1, Sec. 4, Roosevelt Rd, Taipei 106319, Taiwan (R.O.C.)}

\author[0000-0003-4022-4132]{Woojin Kwon}
\affiliation{Department of Earth Science Education, Seoul National University, 1 Gwanak-ro, Gwanak-gu, Seoul 08826, Republic of Korea}
\affiliation{SNU Astronomy Research Center, Seoul National University, 1 Gwanak-ro, Gwanak-gu, Seoul 08826, Republic of Korea}
\affiliation{The Center for Educational Research, Seoul National University, 1 Gwanak-ro, Gwanak-gu, Seoul 08826, Republic of Korea}

\author[0000-0002-3179-6334]{Chang Won Lee}
\affiliation{Korea Astronomy and Space Science Institute (KASI), 776 Daedeokdae-ro, Yuseong-gu, Daejeon 34055, Republic of Korea}
\affiliation{University of Science and Technology, Korea (UST), 217 Gajeong-ro, Yuseong-gu, Daejeon 34113, Republic of Korea}

\author[0000-0003-3119-2087]{Jeong-Eun Lee}
\affiliation{Department of Physics and Astronomy, Seoul National University, 1 Gwanak-ro, Gwanak-gu, Seoul 08826, Korea}

\author[0000-0002-5286-2564]{Tie Liu}
\affiliation{Key Laboratory for Research in Galaxies and Cosmology, Shanghai Astronomical Observatory, Chinese Academy of Sciences, 80 Nandan Road, Shanghai 200030, People’s Republic of China}

\author[0000-0003-4506-3171]{Qiuyi Luo}
\affiliation{Institute of Astronomy, Graduate School of Science, The University of Tokyo, 2-21-1 Osawa, Mitaka, Tokyo 181-0015, Japan}
\affiliation{Department of Astronomy, School of Science, The University of Tokyo, 7-3-1 Hongo, Bunkyo, Tokyo 113-0033, Japan}

\author[0000-0002-7125-7685]{Patricio Sanhueza}
\affiliation{Department of Astronomy, School of Science, The University of Tokyo, 7-3-1 Hongo, Bunkyo, Tokyo 113-0033, Japan}

\author[0000-0001-8385-9838]{Hsien Shang (尚賢)}
\affiliation{Institute of Astronomy and Astrophysics, Academia Sinica, No.1, Sec. 4, Roosevelt Rd, Taipei 106319, Taiwan (R.O.C.)}

\begin{abstract} 
Complex organic molecules (COMs) in young stellar objects (YSOs) have attracted significant attention in recent years due to their potential connection to pre-biotic chemistry and their utility as tracers of warm or shocked gas components. 
Proto-binary and multiple systems with close separations are particularly valuable targets for investigating chemical inheritance and reaction, as their members are expected to form from similar material in their parental cloud. 
We present ALMA observations of the hierarchical proto-triple system HOPS-288, focusing on the physical structure, kinematics, and COM compositions. 
The system is treated as a proto-binary system consisting of HOPS-288-A and HOPS-288-B due to the limited spatial resolutions, with a separation of 200~au. 
Three COM-rich features are revealed: two hot corinos associated with the two members, rich in a variety of COMs, and an intervening component between the two members traced by CH$_3$OH and tentatively by CH$_3$CHO.
The hot corino in HOPS-288-A exhibits rotational features and might trace a disk. 
The hot corino in HOPS-288-B is also possibly exhibiting rotational motion. 
The intervening component could possibly trace a shocked region in the circumbinary disk or a bridge between the two members.
The column densities of COMs, including $^{13}$CH$_3$OH, CH$_2$DOH, CH$_3$CHO, HCOOCH$_3$, C$_2$H$_5$OH, $^{13}$CH$_3$CN, and NH$_2$CHO, are broadly similar between the two sources, possibly suggesting the complex organic similarities among proto-binary/multiple systems.  
Given the complexity of the studied physical structures, further detailed investigations will be essential to confirm this result.
\end{abstract}


\section{Introduction}
\label{sec:Intro}

\begin{deluxetable*}{rrrrrrl}
\label{tab:spw:Tobin}
\caption{
Information of the spectral windows of the high-resolution data (\#2018.1.01038.S) used in this study. 
}
\tablehead{
\colhead{SPW} & \colhead{$f_\mathrm{c}$} & \colhead{BW} & \colhead{d$f$} & \colhead{d$v$} & \colhead{$\sigma$} & \colhead{Molecule} \\
\colhead{} & \colhead{(MHz)} & \colhead{(kHz)} & \colhead{(kHz)} &\colhead{(km s$^{-1}$)} & \colhead{(\mJyperbeam)} & \colhead{}
}
\startdata
\#23 & 219552.6 &   60 &  31 & 0.042 & 7.3 & C$^{18}$O \\ 
\#29 & 218456.3 &   60 & 122 & 0.17 & 3.5 & CH$_3$OH \\   
\#35 & 230529.7 &  938 & 488 & 0.64 & 2.5 & CO, CH$_3$CHO, C$_2$H$_5$OH, and one CH$_3$OH line \\ 
\#37 & 233491.8 & 1875 & 977 & 1.3 & 1.5 & HCOOCH$_3$, C$_2$H$_5$CN, NH$_2$CHO, and two CH$_3$OH lines 
\enddata
\tablecomments{
$f_\mathrm{c}$ is the center frequency. 
BW is the bandwidth. 
d$f$ and d$v$ are the spectral resolution and the corresponding velocity resolution, respectively. 
$\sigma$ is the noise level. 
}
\end{deluxetable*}

Complex organic molecules (COMs) in young stellar objects (YSOs) have attracted significant attention in recent years due to their potential connection to prebiotic chemistry in subsequent planetary systems.
COMs can be formed either on the icy mantles of dust grains and later released via desorption or directly in the gas-phase with previously desorbed parent molecules \citep[see ][and the references therein]{2023Ceccarelli_review}. 
Gas-phase COMs are detected in YSOs at both the starless \citep[e.g.,][]{2020Scibelli_COM_Taurus,2024Scibelli_COM_Perseus,2025Hsu_ALMASOP_starless} and protostellar stages \citep[e.g.,][]{2020Belloche_CALYPSO,2020Hsu_ALMASOP,2021Yang_PEACHES,2022Hsu_ALMASOP,2022Bouvier_ORANGES}.
In particular, the compact, warm regions enriched in COMs surrounding low- and intermediate-mass protostars are referred to as ``hot corinos'' \citep[e.g., ][]{2004Ceccarelli_HotCorino,2018Ospina-Zamudio_CepEmm}. 
Hot corinos are generally attributed to heating of the innermost envelope by the central protostar \citep[e.g.,][]{2004Ceccarelli_HotCorino,2019Jacobsen_L483_COM,2023Hsu_ALMASOP}, although disk-origin hot corinos have also been reported \citep[e.g.,][]{2017Lee_HH212}.
In addition to protostellar heating, shocks can also drive the release of COMs via localized ice sublimation and/or sputtering \citep[e.g.,][]{2002Velusamy_L1157,2008Arce_L1157-B1_COMs,2016Oya_IRAS-16293-2422,2022Okoda_B335_chem,2023Hsu_ALMASOP,2024Hsu_MMS6,2025Bouvier_HOPS409,2025Hsu_G192.12-11.10_COM}.
Because their presence requires efficient desorption mechanisms, COMs serve as valuable tracers of the morphologies and kinematics of either warm or shocked regions in protostellar environments \citep[e.g.,][]{2019Jacobsen_L483_COM,2019Lee_HH212_COM_atm,2025Hsu_G192.12-11.10_COM}.

Several YSOs harboring COMs are simultaneously belonging to binary/multiple systems, such as 
IRAS 16293–2422, \citep[e.g., ][]{2016Jorgensen_PILS,2018Jorgensen_PILS,2020Manigand_IRAS16293_COM}, 
L1551 IRS5 \citep{2020Bianchi_L1551-IRS5_COM}, 
L1448 IRS3B \citep{2021Yang_PEACHES},
Ser-emb 11 \citep{2021Martin_Ser-emb-11_COM}, 
[BHB2007] 11 \citep{2022Vastel_BHB2007-11_COMs}, 
NGC 1333 IRAS 4A \citep{2017Lopez_IRAS4A_COM,2019Sahu_IRAS4A_hot-corino-atm,2020DeSimone_IRAS4A_COM}, 
SVS 13A \citep[e.g.,][]{2019Bianchi_SVS13A_COM,2021Yang_PEACHES,2022Bianchi_SVS13A_COM,2023Hsieh_SVS13A_N,2024Hsieh_SVS13A_COM,2025Hsieh_SVS13A_12C13C}. 
Binary/multiple systems present more complex physical environments than single stars, involving intricate interactions among protostars \citep[e.g., ][]{2012Pineda_IRAS16293_bridge,2019vanderWiel_IRAS16293_bridge}, disks \citep[e.g., ][]{2022Vastel_BHB2007-11_COMs}, and outflows \citep[e.g., ][]{2015Santangelo_IRAS4A_outflow}. 
In recent years, COMs have been recognized as valuable tracers for probing these environments in proto-binary/multiple systems.
For example, in the proto-binary system [BHB2007] 11, \citet{2022Vastel_BHB2007-11_COMs} identified three CH$_3$OH components and proposed that two of them trace localized warm regions in the circumbinary disk, likely heated by mass-accretion streamers. 
Similarly, in the proto-binary system SVS 13A, \citet{2024Hsieh_SVS13A_COM} concluded that COM emission cannot be explained by protostellar radiation alone, and that additional mechanisms, such as accretion shocks induced by large-scale infalling streamers, are likely at play.

Particularly for proto-binary/multiple systems with close separations (a few hundred au), the members are presumed to form from the same parental cloud and thus inherit comparable chemical ingredients.
Accordingly, such systems are expected to host similar chemical compositions.
However, direct comparisons of complex organic compositions within proto-binary/multiple systems remain challenging.
First, for systems having merely a few hundred au, the emissions are difficult to distinguish. 
Meanwhile, the COM emission from YSOs can be weak due to the insufficient luminosity and gas density \citep{2023Hsu_ALMASOP} or be obscured by dust opacity \citep{2019Sahu_IRAS4A_hot-corino-atm,2020DeSimone_IRAS4A_COM}.
Despite these challenges, \citet{2022Bianchi_SVS13A_COM} investigate the COM compositions between the two members of SVS 13 A, which has a separation of 90 au. 
Their evaluated column density ratios between the two sources are around 1 for CH$_3$OH, $^{13}$CH$_3$OH, and CH$_3$OCH3; 2 for CH$_3$CHO; and 4 for NH$_2$CHO. 
They suggested that the observed chemical segregation is likely due to the onion-like distribution of COMs. 
More investigations on the chemical compositions among the proto-binary/multiple systems are still desired. 

HOPS-288 is a protostellar system cataloged by the Herschel Space Observatory \citep{2016Furlan_HOPS} and also observed with a target name of G211.47-19.27S under the project ALMA Survey of Orion PGCCs \citep[ALMASOP; ALMA: Atacama Large Millimeter/sub-millimeter Array; PGCCs: Planck Galactic Cold Clumps; ][]{2020Dutta_ALMASOP} and G211.47-19.27South in surveys such as \citet{2018Yi_PGCC_SCUBA2_II}, \citet{2020Kim_cores_Nobeyama}, and \citet{2021Tatematsu_cores_Nobeyama}. 
Under the investigations of ALMASOP at 1.3 mm (230~GHz), this source was reported to be extremely rich in COMs \citep{2020Hsu_ALMASOP,2022Hsu_ALMASOP}. 
High-resolution imaging from the VLA/ALMA Nascent Disk and Multiplicity (VANDAM) survey \citep{2020Tobin_VANDAM_disk} revealed that HOPS-288 is a binary system at 0.87~mm (345~GHz), HOPS-288-A and HOPS-288-B (hereafter source A and source B, respectively), with a separation of 200~au, and the former itself is a binary system having a separation of 50~au at 9~mm (33~GHz). 

In this paper, we investigate the origin of COM emission in the HOPS-288 system and compare the COM compositions between its members.
The two observational programs used in this study are introduced in \S\ref{sec:Obs}.
In \S\ref{sec:phys}, we examine the overall physical structure of the system, including the outflows and circumbinary disk.
The COM morphologies and kinematics are presented in \S\ref{sec:COM}, providing insights into the origins of the observed COM emission.
In \S\ref{sec:chem}, we derive the column densities and rotational temperatures of several molecules including COMs.
We discuss the implications of our findings in \S\ref{sec:Discussions} and summarize our conclusions in \S\ref{sec:Conclusions}.


\section{Observations}
\label{sec:Obs}

This study makes use of data obtained from two ALMA Cycle~6 programs: \#2018.1.01038.S (PI: John Tobin) and \#2018.1.00302.S (PI: Tie Liu). 
Both of them were observed in Band~6 (1.3~mm, 230~GHz). 
The former provides higher spatial resolution, while the latter covers wider bandwidth.

The high-resolution observation (\#2018.1.01038.S) was carried out with the 12-m array only. 
The unprojected baselines spanned 11.6–1936k$\lambda$, giving a maximum recoverable scale of $\sim$3\farcs7. 
The targeted phase center was $(\alpha_{2000}, \delta_{2000})=$ (05:39:56.0 $-$07:30:27.6), and the absolute flux scale uncertainty is around 10\%. 
The spectral setup included eight SPWs with different resolutions; here we make use of SPWs \#23, \#29, \#35, and \#37, which cover CO, C$^{18}$O, CH$_3$OH, CH$_2$DOH, CH$_3$CHO, HCOOCH$_3$, C$_2$H$_5$OH, and C$_2$H$_5$CN. 
Details of the bandwidths, spectral resolutions, and the corresponding molecular transitions are summarized in Table~\ref{tab:spw:Tobin}.
Imaging was performed with \texttt{tclean} in CASA, with robust = 0.5. 
The synthesized beam was $\sim$0\farcs21 \citep[corresponding to 84~au at a distance of $\sim$400~pc; ][]{2011Lombardi_2MASS_extinction_IV}. 
The achieved sensitivity was $\sim$0.19~\mJyperbeam\ for the continuum (0.1~K at 230~GHz), while the per-channel sensitivities are listed in Table~\ref{tab:spw:Tobin}. 
Owing to its higher spatial and spectral resolutions, this dataset is primarily used to investigate the morphologies and kinematics of selected molecular tracers.
We summarize the transition information in Table~\ref{tab:appx:trans:Tobin}. 

The wide-bandwidth (\#2018.1.00302.S) data employed ALMA the 12-m array. 
The unprojected baselines ranged from 11.6 to 1075~k$\lambda$, corresponding to a maximum recoverable scale of $\sim$3\farcs7. 
The targeted phase center was [05:39:56.1 $-$07:30:28.4], and the absolute flux scale uncertainty is around 10\%. 
The four spectral windows each had a bandwidth of 1,875~MHz and a spectral resolution of 1.129~MHz ($\sim$1.4\kmpers\ at 230GHz). Imaging was performed with \texttt{tclean} in CASA \citep[][]{casa:2022}, adopting a Briggs robust parameter of 0.5.
The resulting synthesized beam was $\sim$0\farcs37 \citep[corresponding to $\sim$148~au at a distance of $\sim$400~pc; ][]{2011Lombardi_2MASS_extinction_IV}. 
The achieved sensitivities were $\sim$0.64\mJyperbeam\ for the full-band continuum (0.13 at 230~GHz) and $\sim$3.6~\mJyperbeam\ per channel.
Given its broad spectral coverage, this dataset is primarily used for deriving chemical compositions.
We summarize the transition information in Table~\ref{tab:appx:trans:almasop}. 

The visibility data of both observations have been applied the self-calibration of phase using the initially exported continuum multi-frequency synthesis image. 
The calibration tables were obtained by \textit{gaincal} with infinite solution interval. 
The calibration were applied to the visibility data by \textit{applycal} with linear interpolation. 
In the high-resolution observations, the dust-continuum peak flux density increased by 50\% and the signal-to-noise ratio (SNR) by a factor of three. 
In the wide-coverage observations, the corresponding increases were 3\% and 25\%.
For the continuum-subtracted data cubes, the line flux density and its SNR in the high-resolution observations improved by around ten percent, whereas the corresponding improvements in the wide-coverage observations were a few percent.

Since neither of the two data sets resolves the local binarity of source A, in this study we treat source A as a single system. 
The position is set at the peak of the unresolved dust continuum at source A. 


\section{Physical Structure of the Multiple System}
\label{sec:phys}

We first investigated the physical structures of the system using the high-resolution (\#2018.1.01038.S) data.


\begin{deluxetable*}{rrrrrrr}
\label{tab:cont}
\caption{\label{tab:fit2D} The 2D Gaussian fitting results of the dust continuum of the two sources. 
}
\tablehead{
\colhead{Source} & \colhead{RA} & \colhead{DEC}    & \colhead{Image Size}           & \colhead{Deconv. size}  & \colhead{$F_\nu$} & \colhead{$I_\nu^{\rm peak}$} \\
\colhead{} & \colhead{(J2000)} & \colhead{(J2000)} & \colhead{(MAJ, MIN, PA)} & \colhead{(MAJ, MIN, PA)} & \colhead{(mJy)} & \colhead{(\mJyperbeam)} 
}
\startdata
HOPS-288-A & 05:39:56.00 & $-$07:30:27.6  & 0\farcs44, 0\farcs25, 141\degr & 0\farcs{40}, 0\farcs{19}, 144\degr & 278 & 78 \\
HOPS-288-B & 05:39:56.03 & $-$07:30:28.0  & 0\farcs34, 0\farcs20, 116\degr & 0\farcs{27}, 0\farcs{13}, 116\degr & 51 & 23 \\
\enddata
\tablecomments{
Deconv. size is the image component size deconvolved by the beam. 
$I_\nu^{\rm peak}$ is the peak specific intensity. 
$F_\nu$ is the flux density. 
}
\end{deluxetable*}

\subsection{Dust}

\begin{figure}[htb!]
\centering
\includegraphics[width=\linewidth]{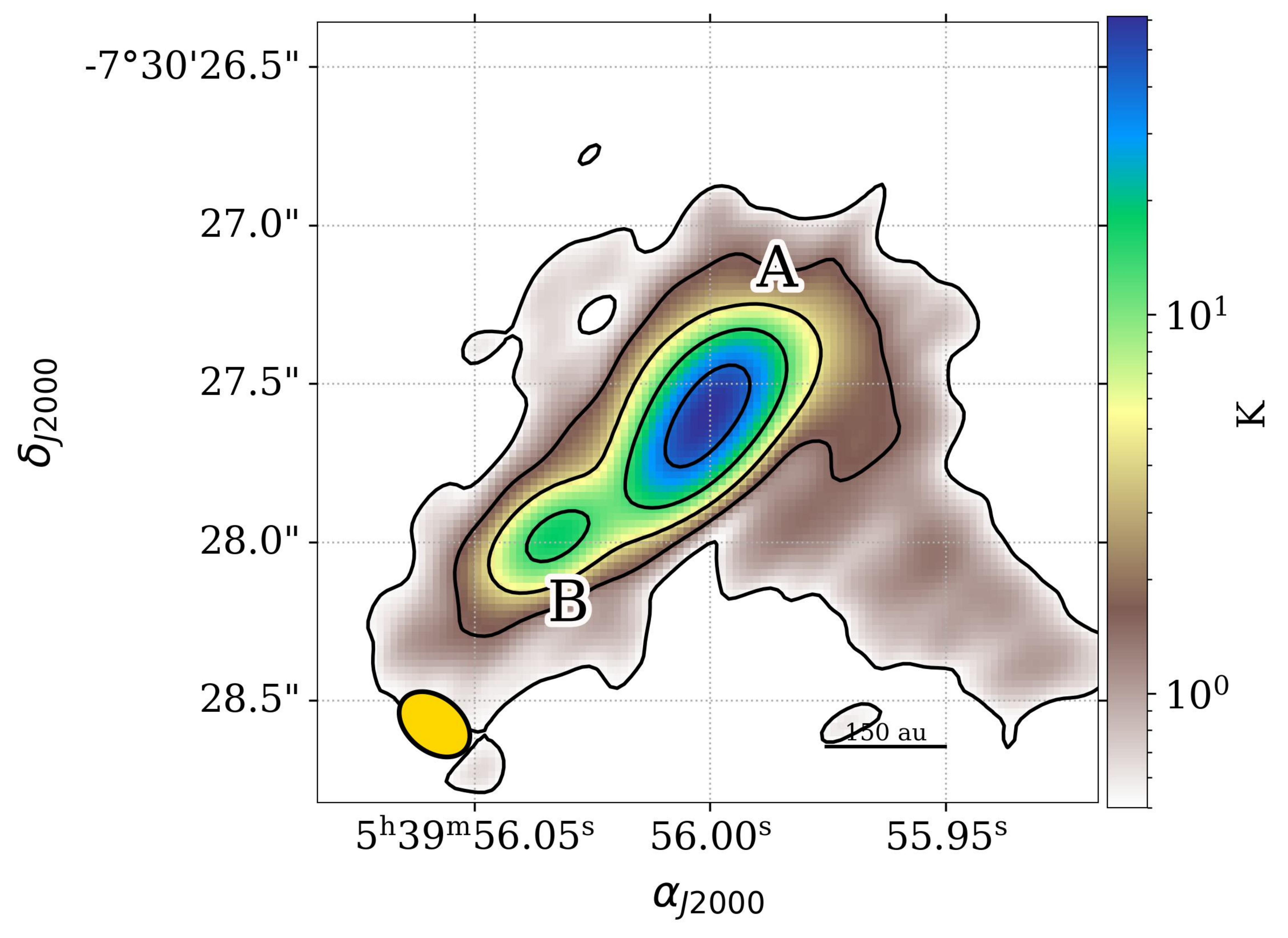}
\caption{\label{fig:img_cont} 
1.3~mm dust continuum observed by the high-resolution data. 
The contour levels are [5, 15, 45, 135, 405]$\sigma$, where $\sigma$ is 0.10 K. 
The yellow ellipse at the bottom left corner illustrates the beam.  
The texts ``A'' and ``B'' label the two continuum peaks. 
}
\end{figure}

Figure~\ref{fig:img_cont} presents the 1.3~mm dust continuum from the high-resolution (\#2018.1.01038.S) data. 
The lowest contour (5$\sigma$) is elongated and encompasses both source A and source B, possibly tracing a circumbinary disk surrounding the system.
Such a flattened morphology suggests that the system may be oriented close to edge-on. 
The third-lowest contour (32$\sigma$) appears to trace the localized structure of each source more distinctly.
At the positions of source A and source B, the continuum emission shows slight elongation, which may correspond to individual circumstellar disks around each source.
The dust continuum emission is seen to extend southwest from the system.
Further self-calibration could improve the sensitivity and potentially reveal this structure more clearly, although such refinements are beyond the scope of the present work.
We applied 2D Gaussian fitting on the dust continuum image and summarize the 2D Gaussian fitting results in Table~\ref{tab:fit2D}. 
For completeness, we show the comparisons of the 1.3~mm dust continuum images from both high-resolution (\#2018.1. 01038.S) and wide-bandwidth (\#2018.1.00302.S) in Figure~\ref{fig:appx_img}. 


\subsection{Outflows}
\label{sec:outflow}

\begin{figure}[htb!]
\centering
\includegraphics[width=1\linewidth]{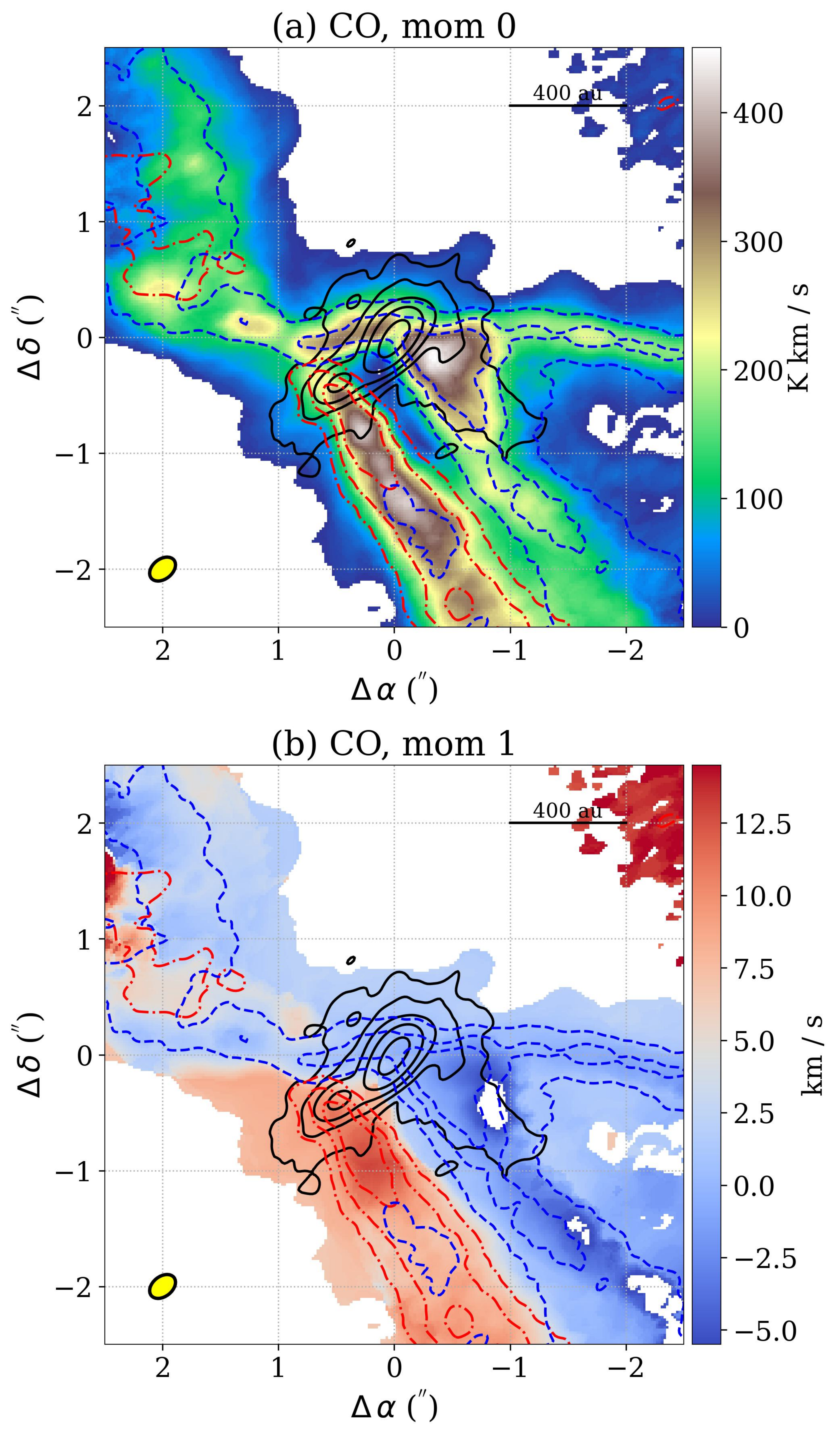}
\caption{\label{fig:img_outflow} 
(a) Integrated intensity and (b) intensity-weighted mean velocity maps of CO $J=2$--$1$, overlaid with 1.3~mm dust continuum (black solid contours) and CO $J=2$--$1$ integrated intensity in red-shifted (red dash-dotted contours) and blue-shifted (blue dashed contours) velocity channels.
All the data were obtained from the high-resolution observations. 
For panel (a), the integration interval was from -15 to 25 \kmpers. 
A mask of 5$\sigma$ was applied during the integration. 
The black contours are at [5, 15, 45, 135, 405]$\sigma$, where $\sigma$ is 0.10 K. 
The velocity intervals of the red dash-dotted and blue dashed contours are [$+10$, $+25$] and [$-15$, $0$] \kmpers, respectively. 
Both are at [10, 30, 50]$\sigma$, where $\sigma$ is 5.64 K km s$^{-1}$. 
The origin of the coordinates is set to the position of source A. 
}
\end{figure}

Figure~\ref{fig:img_outflow} presents the integrated intensity (moment 0) and intensity-weighted mean velocity (moment 1) images of the CO $J=2-1$ transition overlaid with the 1.3~mm dust continuum.
As shown in Figure~\ref{fig:img_outflow} (a), the moment 0 image of CO reveals outflows launched from both source A and source B, particularly visible on the southwestern side of the system.
The orientations of the outflows are consistent with the minor axes (i.e., perpendicular to the major axes) of their originated sources.
Also, the southwestern extension of the dust continuum illustrated by the black solid contours possibly traces the cavity wall of source A's outflow, similar to the case of L1551 IRS 5 \citep{2025Sabatini_L1551IRS5_dustwall}. 

At the southwestern side, the outflow from source A clearly exhibits a wider opening angle than that from source B. 
Due to the slight difference in axis orientation between the two sources, the outflows overlap in the northeastern region, making the outflow lobes there hardly distinguished. 
To better distinguish the outflow lobes in the northeastern region, we plot the CO $J=2–1$ integrated intensity contours for the red- and blue-shifted channels separately. 
As shown in Figure~\ref{fig:img_outflow}(a), the cavity wall of source A’s outflow can be clearly identified in the northeastern region, whereas the cavity of source B’s outflow remains unclear.

The different opening angles may suggest that source A is at a more advanced evolutionary stage \citep[e.g.,][]{2023Hsieh_outflow,2024Dunham_outflow}.
Alternative explanations for the wide-opening morphology of source A, considering the binarity of source A, include the presence of two outflows \citep[e.g., NGC 1333 IRAS 4A;][]{2015Santangelo_IRAS4A_outflow} or jet precession \citep[e.g.,][]{2009Raga_jet_precession,2018Louvet_HH30}. 
Additional support for an evolutionary difference between source A and source B comes from the (non-)detection of collimated SiO jets.
SiO jets are typically observed in Class 0 rather than Class I protostars, though a few detections in Class I objects have been suggested \citep{2022Dutta_ALMASOP_SiO}. 
\citet{2024Dutta_ALMASOP_jet} detect a collimated SiO $J=5$--$4$ jet only from source B (see also Figure~\ref{fig:appx_img}). 
This supports that source A is more evolved, whereas a possible weak jet indicated by the compact SiO emission detected at source A remains unconfirmed.

Finally, as shown in Figure~\ref{fig:img_outflow}(b), the outflows from source A and source B are predominantly blueshifted and redshifted, respectively.
The relatively uniform speed within each outflow suggests that they are viewed close to edge-on.


\subsection{Circumbinary Disk}
\label{sec:CBD}

\begin{figure}[htb!]
\centering
\includegraphics[width=1\linewidth]{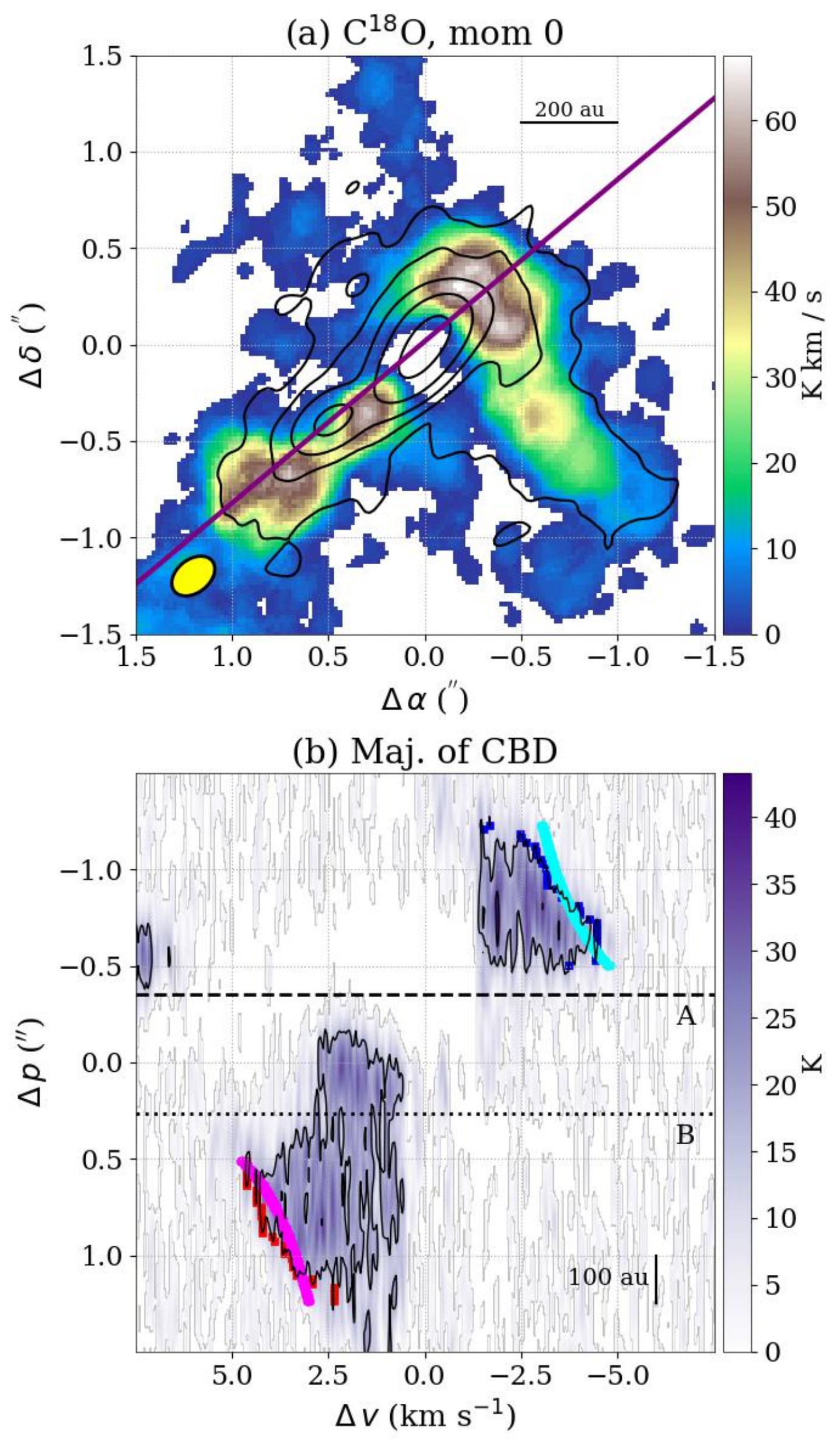}
\caption{\label{fig:img_CBD} 
Integrated intensity image (a) and PV diagram (b) of the C$^{18}$O $J=2-1$ transition from high-resolution observations. 
The black contours in (a) show the 1.3 mm continuum at [5, 15, 45, 135, 405]$\sigma$. 
Coordinates in panel (a) is centered on source A. 
A mask of 5$\sigma$ was applied during the integration. 
The PV diagram (b) is extracted along the circumbinary disk axis illustrated by the purple line in panel (a).
The black contour in (b) shows the 5$\sigma$ of C$^{18}$O emission. 
The reference velocity ($\Delta v=0$ \kmpers) of the PV diagram is set to 5 km s$^{-1}$. 
The reference position ($\Delta p = 0$ \kmpers) was set at the gap between source A and source B (5:39:56.02, $-$7:30:27.8). 
The red and blue markers represent the datapoints used for fitting the rotation motion of the C$^{18}$O gas component. 
The magenta and cyan markers show the modeling results. 
}
\end{figure}

Figure~\ref{fig:img_CBD}(a) presents the integrated intensity map of C$^{18}$O (color scale) overlaid with the 1.3~mm dust continuum (black contours).
The C$^{18}$O emission shows three bright components distributed from southeast to northwest: (i) one covering the southeast side of the dust continuum, (ii) one located in the gap between source A and source B, and (iii) one covering the northwest side of the dust continuum, which, similar to the dust, extends toward the southwest. 
The morphology, with C$^{18}$O lying along the midplane and flanking both sides of the elongated dust continuum, suggests that the two C$^{18}$O component (i) and (iii) are possibly tracing the circumbinary disk.

To model the possible disk motion, we extracted a position–velocity (PV) diagram along the cut penetrating source A and B (PA = $127^{\circ}$), as indicated by the purple line in Figure \ref{fig:img_CBD}(a). 
The resulting PV diagram is shown in Figure~\ref{fig:img_CBD}(b), with the reference position set at the gap between source A and source B. 
The reference velocity, defined as the system velocity of the circumbinary disk $v_\mathrm{CBD}$, was set at 5.0 km s$^{-1}$ based on the best symmetry of the PV diagram. 

The PV diagram shown in \ref{fig:img_CBD}(b) reveals three features located at $\Delta p = +0\farcs75$, $0\farcs0$, and $-0\farcs75$, corresponding to the three bright components from southeast to northwest in Figure~\ref{fig:img_CBD}(a). 
The features at $\Delta p = +0\farcs75$ and $-0\farcs75$ correspond to the red- and blueshifted sides of the circumbinary disk, respectively.
Note that the red- and blueshifted motion are consistent with that observed in the CO outflow in Figure~\ref{fig:img_outflow}(b). 

We assumed that the circumbinary disk follows Keplerian rotation around a central mass ($M_\mathrm{cen}$) as $v = \left ( GM_\mathrm{cen}/R \right )^{1/2}$, where $R$ is the radius. 
Based on the spatial extent of the blueshifted component at $\Delta p = -0\farcs75$, we adopt an outer radius of $1\farcs25$ (500~au) for the circumbinary disk.
We constructed a grid of parameters by varying the central mass ($M_\mathrm{cen}$) from 1 to 10~\Msun\ in steps of 0.25~\Msun.
We also varied the inclination angle ($\varphi$), defined such that $90^\circ$ corresponds to a perfectly edge-on geometry, between $75^\circ$ and $105^\circ$ in steps of $5^\circ$.
The fitting was evaluated by the sum of squared error between the relative velocity ($\Delta v$) at each relative position ($\Delta p$). 
The observed velocity was evaluated by the maximum value of $5\sigma$, as indicated by the red and blue markers in Figure~\ref{fig:img_CBD}(b). 

The best-fit central mass ($M_\mathrm{cen}$) is 5.00~\Msun\ for an edge-on orientation ($\varphi = 90^\circ$),  5.25~\Msun\ for a tilt of $5^\circ$ and $10^\circ$, and 5.50~\Msun\ for a tilt of $15^\circ$.
The best-fit velocities at different inclination angles are shown by the cyan and magenta markers in Figure~\ref{fig:img_CBD}(b). 
All of these best-fit models can describe the observations well. 
The relatively high mass suggests that this system is an intermediate-mass protostellar system, which is broadly consistent with its reported high luminosity \citep[$180\pm70$\Lsun,][]{2020Dutta_ALMASOP,2022Hsu_ALMASOP}, though we note that luminosity can also have an accretion component.


\begin{figure*}[htb!]
\centering
\includegraphics[width=\linewidth]{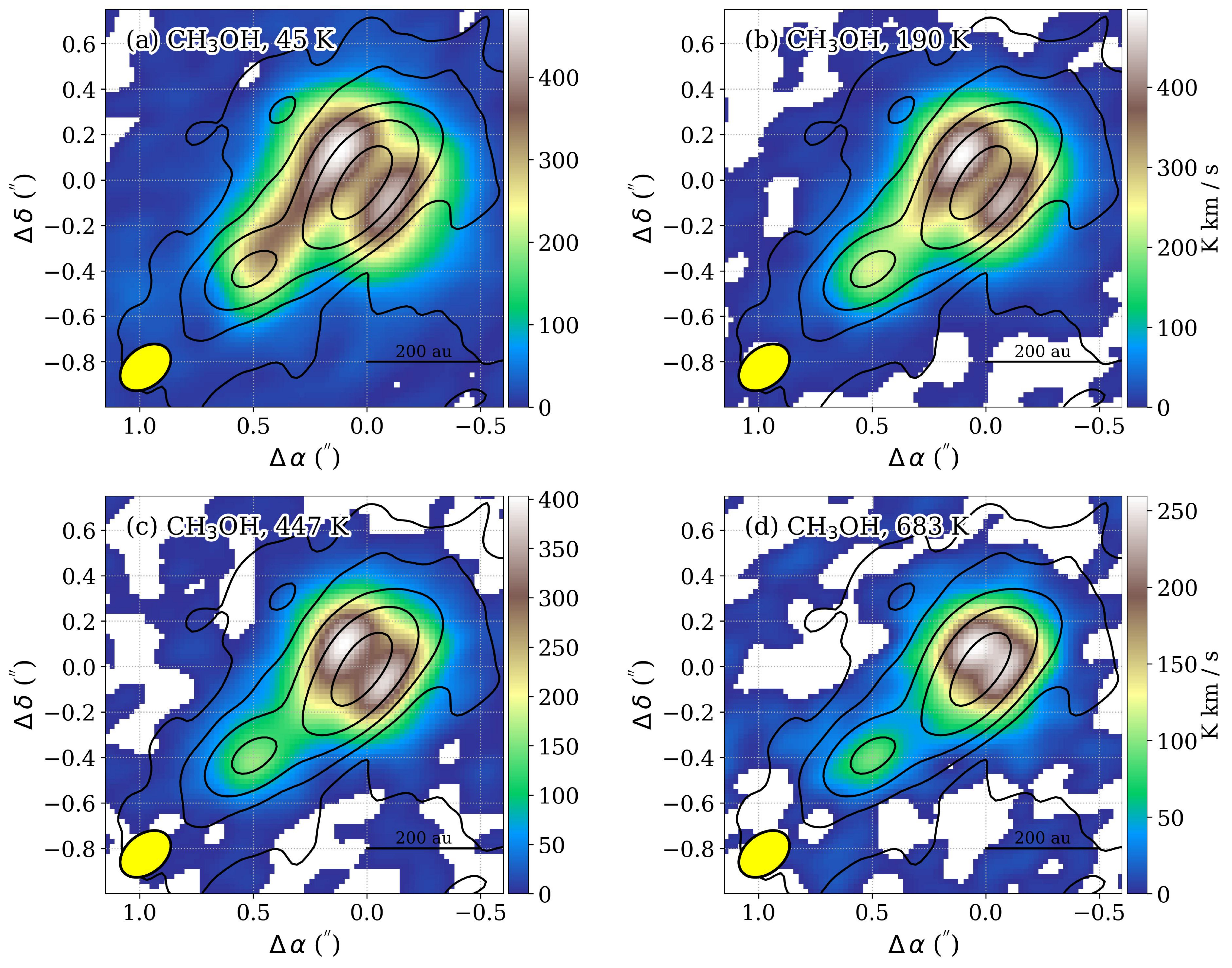}
\caption{\label{fig:mom0_CH3OH} 
The integrated intensity images of CH$_3$OH transitions overlaid with dust continuum (black contours at [5, 15, 45, 135, 405]$\sigma$ and $\sigma=0.10$ K). 
The integration intervals span from $-2.5$ to $+12.5$~\kmpers.
}
\end{figure*}

\section{Morphology and Kinematics of Complex Organics}
\label{sec:COM}

\subsection{Morphology of Methanol}
\label{sec:CH3OH:mom0}

CH$_3$OH is the simplest and most commonly detected COM and is frequently used as a tracer of complex organic chemistry.
Figures~\ref{fig:mom0_CH3OH}(a)–(d) show CH$_3$OH transitions with upper energy levels ranging from \Eu{} = 45 K to 683 K, revealing similar structure with apparent variations.
At \Eu{} = 45 K, Figure~\ref{fig:mom0_CH3OH}(a), the morphology exhibits a hamburger-like morphology covering the dense continuum structure. 
In addition, we identify an intervening component that appears to form a tail connecting the two sources.
The hamburger- and tail-like morphologies gradually become unclear with increasing upper energy levels, as seen by comparing Figure~\ref{fig:mom0_CH3OH}(a) to (d).
At $E_u = 683$ K, Figure~\ref{fig:mom0_CH3OH}(d), two compact CH$_3$OH components become prominent.
The two components locate at the positions of source A and source B, suggesting that both sources host their own hot corinos. 

The hot corino associated with source~A exhibits a dark lane sandwiched by two bright features in Figures~\ref{fig:mom0_CH3OH}(a)–(d), forming a hamburgur-like morphology.
The dark lane likely arises from the midplane material that has relatively low temperature and high optical depth. 
In addition, the northeastern layer is brighter than the southwestern one.
Given that the system is likely viewed near edge-on, the CH$_3$OH emission may originate from the upper and lower surfaces of a disk. 
The emission is brighter in the northeast likely because the disk is tilted slightly to the southwest, allowing us to observe the warmer inner region of the upper surface.
This could be similar to the disk atmosphere case of HH 212 \citep{2017Lee_HH212, 2019Lee_HH212_COM_atm}. 
To confirm this as well as the origin of the unresolved hot corino in source B, we investigate the kinematics of CH$_3$OH in the next section (Section~\ref{sec:CH3OH:PV}). 
We note that the hot corino in source A shows a small offset (0\farcs05) toward the northwest from its continuum peak, indicating an asymmetric structure within source A. 


The intervening component that appears as a tail connecting the two sources in Figure~\ref{fig:mom0_CH3OH}(a) could represent a physical bridge between them, although we cannot rule out the possibility that this structure is an artifact caused by beam convolution.
If the tail-like morphology is indeed artificial, the presence of the intervening emission would instead suggest a localized warm or shocked region in the vicinity of source B.
The absence of this component in the higher–$E_u$ transition further indicates that it is cooler than the two individual hot corinos.

\subsection{Kinematics Revealed by Methanol}
\label{sec:CH3OH:PV}

\begin{figure}[htb!]
\centering
\includegraphics[width=\linewidth]{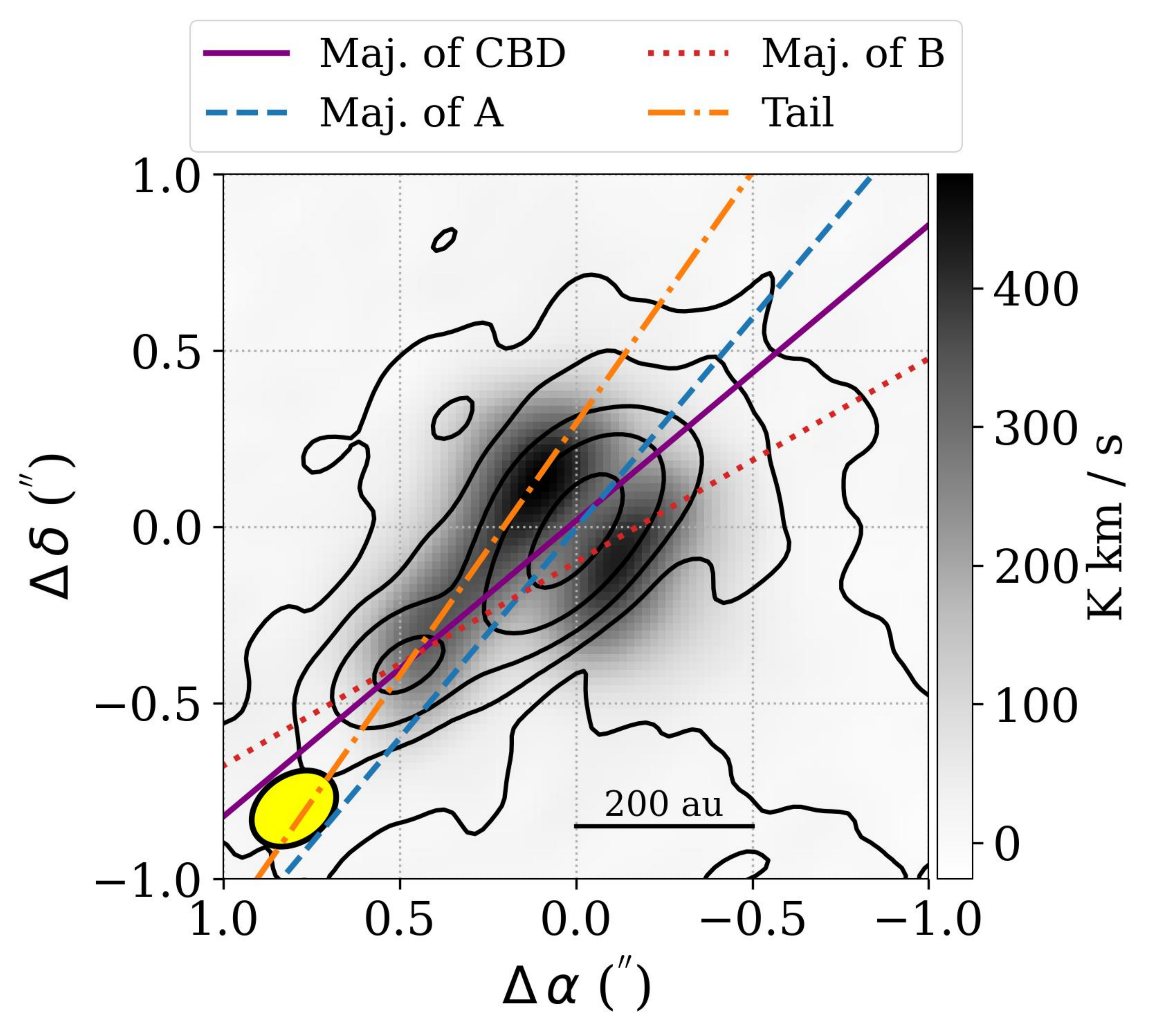}
\caption{\label{fig:img_cuts} 
The cuts used in this study overlaid on integrated intensity image  of CH$_3$OH $4_{2,3}-3_{1,2}$ (\Eu = 45 K) transition (color scale). 
The image is also overlaid by 1.3~mm continuum (black contours at [5, 15, 45, 135, 405]$\sigma$ and $\sigma=0.10$ K). 
}
\end{figure}

\begin{figure*}[htb!]
\centering
\includegraphics[width=\linewidth]{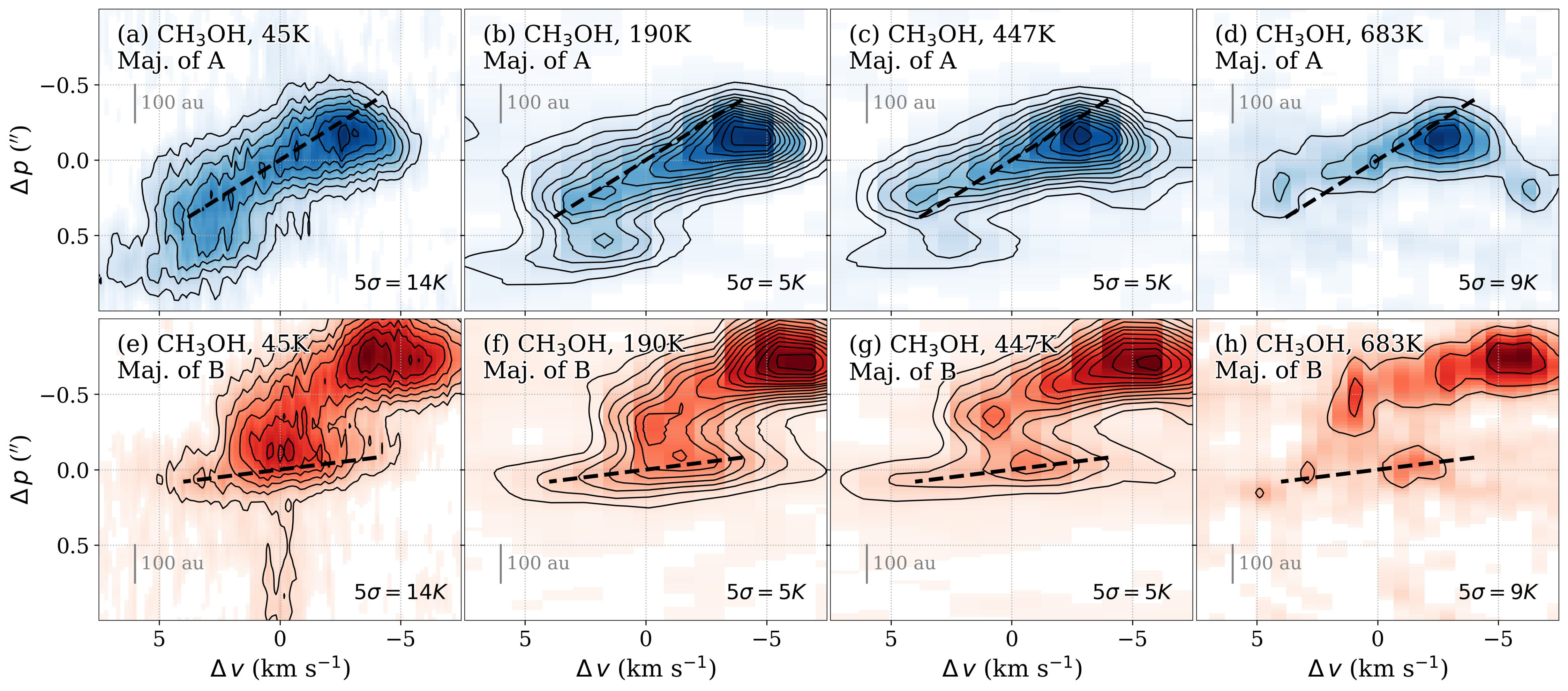}
\caption{\label{fig:PV_CH3OH} 
PV diagrams along the major axes of source A and source B. 
Contours start at 5$\sigma$ and increase in steps of 5$\sigma$.
The reference velocities ($\Delta v = 0$ \kmpers) were set to 3.5 and 6.5 \kmpers\ for source A and source B, respectively, chosen to maximize the symmetry of the diagrams. 
The reference positions ($\Delta p = 0$ \kmpers) were set the the positions of source A and source B for the top and bottom rows, respectively. 
The black dashed lines illustrate the velocity gradient relations derived from panel (a) and (e) for the top and bottom rows, respectively.
}
\end{figure*}

To investigate the kinematics of the COM-rich gas components, we extracted PV diagrams of CH$_3$OH along the major axis of each source.
The cuts (``Maj. of A'' and ``Maj. of B'') are illustrated in Figure~\ref{fig:img_cuts} and the PV diagrams are presented in Figure~\ref{fig:PV_CH3OH}. 
Two additional cuts, ``Maj. of CBD'' and ``Tail,'' indicated in Figure \ref{fig:img_cuts}, are also used for the analyses discussed in the following sections. 
The reference velocities of the PV diagrams ($\Delta v = 0$ \kmpers) for each cut were set to maximize symmetry based on the lowest upper energy transitions, namely Figure~\ref{fig:PV_CH3OH}(a) and (e) for source A and source B, respectively. 
The values are 3.5 and 6.5 \kmpers\ for source A and source B, respectively. 
We chose these transitions because they have the best spectral resolution and the strongest line intensities among our data.
The reference positions correspond to the continuum peak of each source. 

Figure~\ref{fig:PV_CH3OH}(a) presents the PV diagram along the major axis of source A for the CH$_3$OH \Eu$=45$ K transition.
A clear linear correlation between $\Delta p$ and $\Delta v$, indicated by the black dashed line selected by visual inspection, suggests that the gas traces a localized rotational structure, possibly a ring or disk. 
In the upper-right quadrant ($\Delta v < 0$ km s$^{-1}$, $\Delta p < 0\arcsec$), the blueshifted emission exhibits a characteristic of disk motion \citep[e.g.,][]{2017Lee_HH212}.
However, the corresponding redshifted component on the opposite side, expected at $0\arcsec < \Delta p < 0\farcs5$, is absent in Figure~\ref{fig:PV_CH3OH}(a).
As shown in Figure~\ref{fig:PV_CH3OH} (c) and (d), the redshifted component gradually becomes visible at $\Delta p=0\farcs2$ and $\Delta v=4$~\kmpers.
At higher-\Eu\ transitions the redshifted component remains much fainter than the blueshifted side.
This discrepancy is consistent with the offset we observed in the morphology, Figure~\ref{fig:mom0_CH3OH} and described in Section~\ref{sec:CH3OH:mom0}). 

The second row of Figure~\ref{fig:PV_CH3OH} presents the PV diagrams along the major axis of source B for the four CH$_3$OH transitions. 
In Figure~\ref{fig:PV_CH3OH}(e), the black dashed line highlights a tentative linear correlation, indicative of rotational motion in the CH$_3$OH gas. 
Given the limited spatial resolution, we cannot determine whether this structure corresponds to an unresolved ring or disk. 
An offset peak appears at $\Delta v = +0.00$ km s$^{-1}$ and $\Delta p = -0\farcs1$, which is likely associated with the intervening component, as it disappears in the higher-\Eu{} transitions shown in Figures~\ref{fig:PV_CH3OH}(f), (g), and (h). 
As this component diminishes, the hot corino emission in source B becomes increasingly apparent. 
Its peak position, at $\Delta v = -1$ km s$^{-1}$ and $\Delta p = -0\farcs1$, is slightly blueshifted and offset from the continuum peak ($\Delta p = 0\arcs$). 

We further model the rotation motion for source A and source B, which are based on the black dashed lines in Figures~\ref{fig:PV_CH3OH}(a) and (e).  
The black dashed lines, selected by visual inspection, are intended to illustrate the linearity.
The linearity between $\Delta p$ and $\Delta v$ can be described as: 
\begin{equation}
    \label{eq:dp_dv_A}
    \left ( \frac{\Delta p}{0.50~\mathrm{arcsec}} \right )=\left ( \frac{\Delta v}{5~\mathrm{km~s^{-1}}} \right )
\end{equation}
for source A and 
\begin{equation}
    \label{eq:dp_dv_B}
    \left ( \frac{\Delta p}{0.10~\mathrm{arcsec}} \right )=\left ( \frac{\Delta v}{5~\mathrm{km~s^{-1}}} \right )
\end{equation}
for source B. 
Assuming that the disk mass is much smaller than the stellar mass and an edge-on geometry, the relative velocity along the line of sight ($\Delta v$) at a relative position ($\Delta p$) following of a Keplerian rotation is:
\begin{equation}
    \Delta v=\sqrt{\frac{GM_\star}{R}\,\frac{\Delta p}{R}}, 
\end{equation}
where \Mstar\ and $R$ are the stellar mass and rotation radius, respectively. 
Given a distance of 400~pc \citep{2011Lombardi_2MASS_extinction_IV} and adopting the outer boundaries of the PV diagrams as the sizes of the rotating features, 200~au (0\farcs50) for source A and 40~au (0\farcs1) for source B, the inferred stellar masses are 2.9~\Msun\ for source A and 1.1~\Msun\ for source B.
The sum of the two stellar masses (4.0~\Msun) is comparable to the value ($\sim$5.00~\Msun) estimated from the circumbinary disk motion traced by C$^{18}$O in Section~\ref{sec:CBD}. 


\begin{figure}[htb!]
\centering
\includegraphics[width=\linewidth]{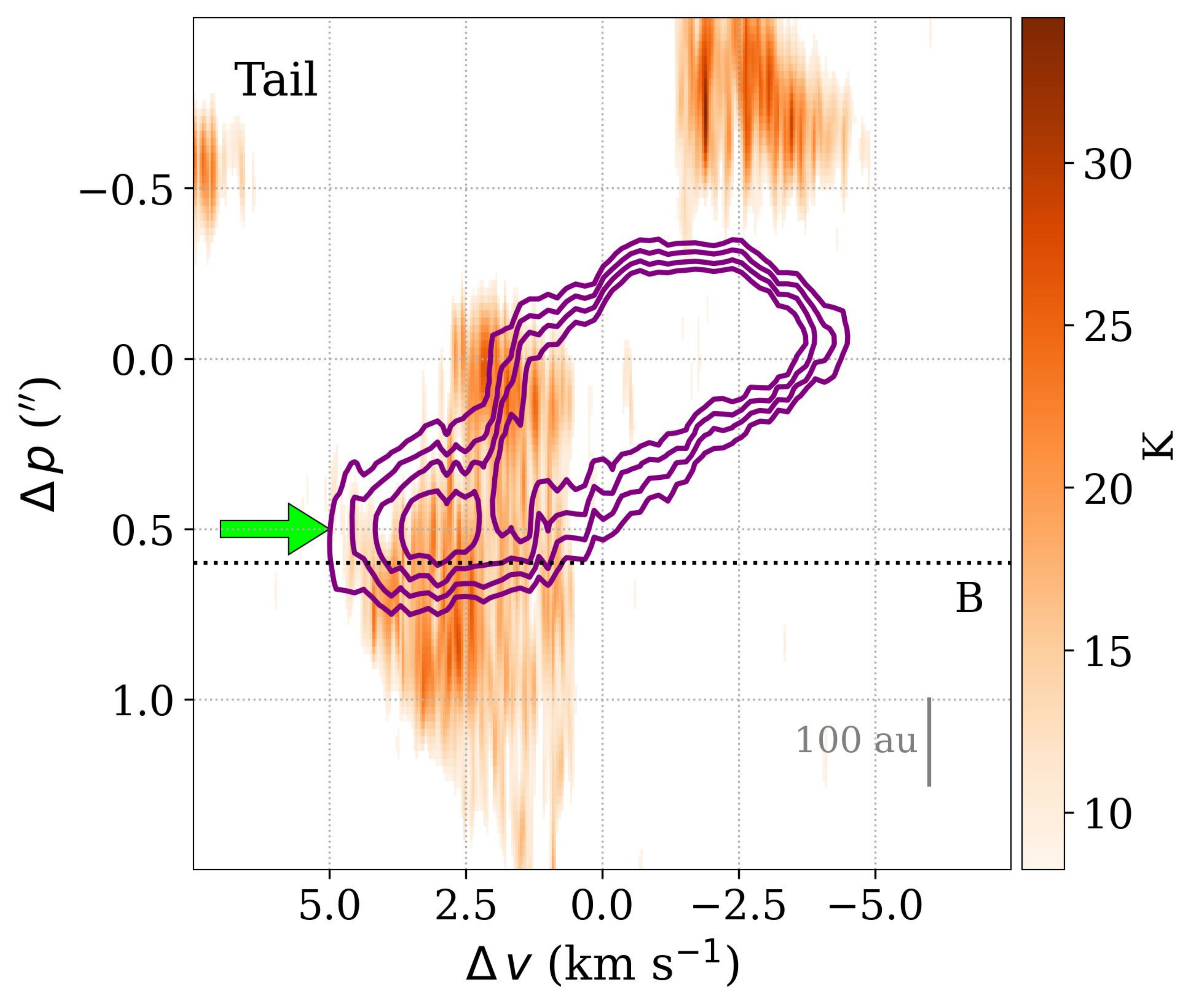}
\caption{\label{fig:PV_tail} 
PV diagrams of the C$^{18}$O (color scale) and CH$_3$OH 45 K transitions (contours) extracted along the tail-like feature. 
The reference velocity ($\Delta v=0$ \kmpers) is set to 5 km s$^{-1}$. 
The reference position ($\Delta p = 0$ \kmpers) is set the the position of source A. 
Contours expand between 20$\sigma$ and 35$\sigma$ in steps of 5$\sigma$.
The green arrow points out the localized enhancement. 
}
\end{figure}

Regarding the kinematics of the intervening component, Figure~\ref{fig:PV_tail} shows the PV diagrams of C$^{18}$O (color scale) and CH$_3$OH 45 K transitions (contours) along the tail. 
The contour levels start from 20$\sigma$, rather than 5$\sigma$, to better visualize this component. 
The cut passes through the source B continuum peak, as illustrated by the orange dash-dotted line in Figure~\ref{fig:img_cuts}. 
As shown in Figure~\ref{fig:PV_tail}, a localized enhancement, previously observed in Figure~\ref{fig:PV_CH3OH}(e), is present at $\Delta p = 0\farcs5$ (green arrow). 
This component appears to coincide with the inner part of the redshifted portion of the circumbinary disk traced by C$^{18}$O, as suggested by the location and velocity of this component. 
This may indicate a localized heated and/or shocked region in the circumbinary disk. 
The potential origins of such a feature include mass flow from source A to source B or an accretion streamer from the surrounding medium. 
A detailed modeling that incorporates accretion flows will be necessary for further confirmation.


\begin{figure*}[htb!]
\centering
\includegraphics[width=\linewidth]{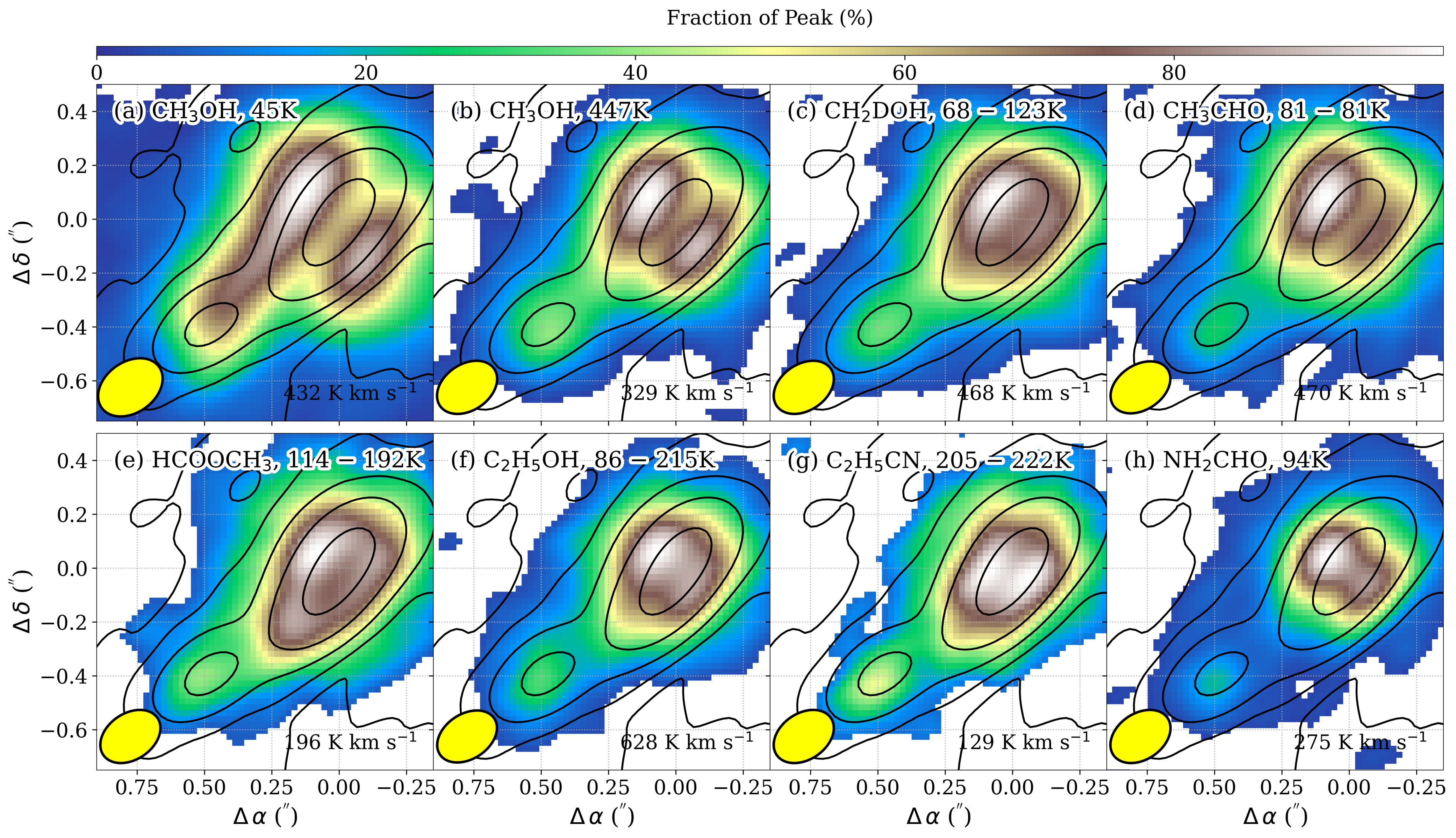}
\caption{\label{fig:mom0_COM} 
The integrated intensity images of selected COM transitions (color scale) overlaid with continuum (black contours at [5, 15, 45, 135, 405]$\sigma$ and $\sigma=0.10$ K). 
The velocity range for integration is fixed to [0, 10] km s$^{-1}$.
}
\end{figure*}

\begin{figure*}[htb!]
\centering
\includegraphics[width=\linewidth]{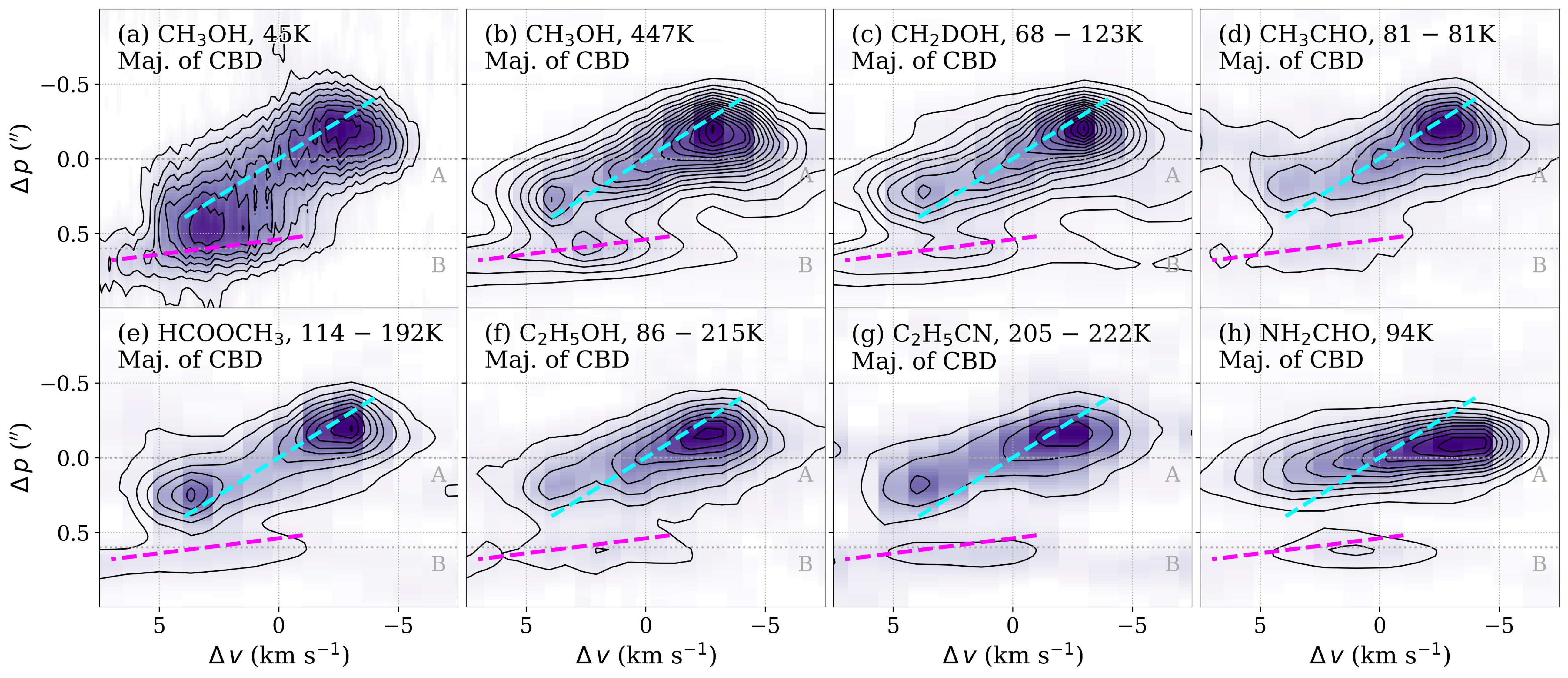}
\caption{\label{fig:PV_COM_Cmajor} 
The PV diagrams cut along the major axis of the circumbinary disk of selected COM transitions. 
The reference velocity is set to 5 km s$^{-1}$. 
The reference position is fixed at the location of source A, while source B appears at $\Delta p = 0\farcs6$, as illustrated by the grey dotted lines.
Contours start at 5$\sigma$ and increase in steps of 5$\sigma$.
Cyan and magenta dashed lines indicate the best-fit velocity structures for source A and source B, respectively, obtained from the CH$_3$OH PV diagrams of the two sources with the methods described in Section~\ref{sec:CH3OH:PV}. 
}
\end{figure*}

\subsection{Diversities and Similarities between Complex Organics}
\label{sec:COM:COMs}

Similar morphologies and kinematics are seen in other COMs, although some differences between those of COMs are present.
In Figure~\ref{fig:mom0_COM}, we present integrated intensity maps for selected COM transitions, while Figure~\ref{fig:PV_COM_Cmajor} shows the corresponding PV diagrams taken along the major axis of the circumbinary disk.
For CH$_2$DOH, CH$_3$CHO, HCOOCH$_3$, C$_2$H$_5$OH, C$_2$H$_5$CN, multiple transitions were stacked to improve the signal-to-noise ratio (SNR).
In the PV diagrams (Figure~\ref{fig:PV_COM_Cmajor}), the reference velocity ($\Delta v=0$ \kmpers) is set to 5 km s$^{-1}$, corresponding to the systemic velocity derived from the rotation of the circumbinary disk.
The reference position is fixed at the location of source A, while source B appears at $\Delta p = 0\farcs6$.
Cyan and magenta dashed lines indicate the best-fit velocity structures for source A and source B, respectively, obtained from the CH$_3$OH PV diagrams with the methods described Section~\ref{sec:CH3OH:PV}.  
Since the major axis of the circumbinary disk differs only slightly ($10^{\circ}$) from those of source A and source B, these modeled correlations (Equations~\ref{eq:dp_dv_A} and \ref{eq:dp_dv_B}) remain robust and provide useful insights on the overall kinematics.
For comparison, panels (a) and (b) in both Figure~\ref{fig:mom0_COM} and Figure~\ref{fig:PV_COM_Cmajor} present the CH$_3$OH transitions with \Eu\ = 45 K and 447 K, respectively.

The morphologies of CH$_2$DOH, CH$_3$CHO, HCOOCH$_3$, C$_2$H$_5$OH, C$_2$H$_5$CN, and NH$_2$CHO all exhibit two distinct hot corinos, as shown in Figures \ref{fig:mom0_COM}(c)–(h).
The tail structure is generally absent, except for CH$_3$CHO, where the tail feature tentatively appears at the top atmosphere in Figure \ref{fig:mom0_COM}(d). 
This suggests that, for these COMs, the observed emission is dominated by the individual hot corinos of the two sources.
Consistently, as shown in Figures \ref{fig:PV_COM_Cmajor}(c)–(g), the PV diagrams of CH$_2$DOH, CH$_3$CHO, HCOOCH$_3$, C$_2$H$_5$OH, and C$_2$H$_5$CN (i.e., all the other COMs excepting NH$_2$CHO) are similar with that of CH$_3$OH 447 K in \ref{fig:PV_COM_Cmajor}(b). 

One exception is NH$_2$CHO. 
As shown in Figure~\ref{fig:PV_COM_Cmajor}(h), the diagram does not follow the typical linear relation between $\Delta p$ and $\Delta v$ observed for the other COMs.
However, based on the contour shapes and the relative $\Delta p$ positions of the intensity peaks, we can still identify that the blueshifted and redshifted components are slightly offset toward the northwestern and southeastern sides of the continuum peak ($\Delta p=0\arcs$), respectively, consistent with the overall rotation of the system.
These characteristics suggest that NH$_2$CHO resides in the more inner part (0\farcs1 in Figure~\ref{fig:PV_COM_Cmajor}(h)) of source A, which is further supported by its limited spatial extent that less extended (0\farcs2 in Figure~\ref{fig:mom0_COM}(h)) along the major axis of source A=. 
See Section~\ref{sec:disc:formamide} for further discussion of the distribution and formation of this species. 

Finally, we noticed that, in Figure~\ref{fig:PV_COM_Cmajor}(e), the redshifted portion of the disk is relatively prominent in HCOOCH$_3$. 
This is consistent with its morphology, as shown in Figure~\ref{fig:mom0_COM}(e), that the southeastern edge of source A is locally enhanced. 



\begin{deluxetable}{lcccc}
\tablefontsize{\scriptsize}
\tablecaption{\label{tab:molec} Evaluated excitation temperatures and total column densities. }
\tablehead{\colhead{Species} & \colhead{$N_\mathrm{tot}^A$} & \colhead{$N_\mathrm{tot}^B$} & \colhead{$T_\mathrm{ex}^A$} & \colhead{$T_\mathrm{ex}^B$} \\
\colhead{} & \colhead{(cm$^{-2}$)} & \colhead{(cm$^{-2}$)} & \colhead{(K)} & \colhead{(K)} 
}
\startdata
DCN;v=0; & 3.6$^{+0.5}_{-0.4}\times$ 10$^{15}$ & 9.7$^{+0.8}_{-1.1}\times$ 10$^{14}$ & 200$^\dagger$ & 200$^\dagger$ \\
H$_2$S;v=0; & 8.3$^{+1.3}_{-1.5}\times$ 10$^{16}$ & 3.2$^{+0.4}_{-0.2}\times$ 10$^{16}$ & 200$^\dagger$ & 200$^\dagger$ \\
OCS;v=0; & 5.3$^{+1.1}_{-1.1}\times$ 10$^{16}$ & 3.0$^{+0.3}_{-0.3}\times$ 10$^{16}$ & 203$^{+18}_{-10}$ & 87$^{+11}_{-9}$ \\
H$_2$CO;v=0; & 2.0$^{+0.4}_{-0.5}\times$ 10$^{17}$ & 1.2$^{+0.1}_{-0.1}\times$ 10$^{17}$ & 187$^{+10}_{-10}$ & 326$^{+11}_{-11}$ \\
D$_2$CO;v=0; & 4.2$^{+0.3}_{-0.3}\times$ 10$^{16}$ & 9.9$^{+0.9}_{-1.3}\times$ 10$^{15}$ & 311$^{+12}_{-9}$ & 333$^{+11}_{-11}$ \\
HNCO;v=0; & 6.1$^{+0.7}_{-0.4}\times$ 10$^{16}$ & 7.5$^{+1.5}_{-0.9}\times$ 10$^{15}$ & 336$^{+12}_{-11}$ & 300$^{+10}_{-15}$ \\
HCCCN;v=0; & 1.5$^{+0.2}_{-0.1}\times$ 10$^{15}$ & 3.3$^{+0.5}_{-0.6}\times$ 10$^{14}$ & 200$^\dagger$ & 200$^\dagger$ \\
CH$_3$OH;v=0; & 2.3$^{+0.4}_{-0.3}\times$ 10$^{18}$ & 8.4$^{+0.4}_{-0.3}\times$ 10$^{17}$ & 218$^{+11}_{-9}$ & 227$^{+8}_{-7}$ \\
CH$_3$OH;v12=1; & 2.7$^{+0.7}_{-0.4}\times$ 10$^{18}$ & 5.5$^{+0.9}_{-0.5}\times$ 10$^{17}$ & 264$^{+9}_{-7}$ & 200$^\dagger$ \\
$^{13}$CH$_3$OH;v=0; & 2.9$^{+0.6}_{-0.4}\times$ 10$^{17}$ & 6.0$^{+1.7}_{-0.8}\times$ 10$^{16}$ & 205$^{+8}_{-8}$ & 178$^{+18}_{-10}$ \\
CH$_2$DOH;v=0; & 1.0$^{+0.2}_{-0.1}\times$ 10$^{18}$ & 1.0$^{+0.3}_{-0.1}\times$ 10$^{17}$ & 119$^{+6}_{-8}$ & 122$^{+10}_{-16}$ \\
$^{13}$CH$_3$CN;v=0; & 2.3$^{+0.3}_{-0.2}\times$ 10$^{15}$ & 6.1$^{+1.1}_{-1.1}\times$ 10$^{14}$ & 297$^{+12}_{-10}$ & 309$^{+9}_{-10}$ \\
CH$_3$CHO;v=0; & 1.2$^{+0.1}_{-0.0}\times$ 10$^{17}$ & 2.1$^{+0.2}_{-0.2}\times$ 10$^{16}$ & 214$^{+10}_{-7}$ & 296$^{+11}_{-17}$ \\
CH$_3$CHO;v15=1; & 1.3$^{+0.1}_{-0.1}\times$ 10$^{17}$ & \nodata & 252$^{+8}_{-8}$ & \nodata \\
NH$_2$CHO;v=0; & 2.2$^{+0.4}_{-0.3}\times$ 10$^{16}$ & 3.3$^{+0.7}_{-0.3}\times$ 10$^{15}$ & 261$^{+14}_{-12}$ & 317$^{+9}_{-13}$ \\
CH$_2$(OH)CHO;v=0; & 6.4$^{+0.6}_{-0.6}\times$ 10$^{16}$ & \nodata & 284$^{+12}_{-10}$ & \nodata \\
HCOOCH$_3$;v=0; & 5.1$^{+0.4}_{-0.4}\times$ 10$^{17}$ & 6.3$^{+1.0}_{-0.5}\times$ 10$^{16}$ & 291$^{+11}_{-11}$ & 262$^{+17}_{-14}$ \\
HCOOCH$_3$;v18=1; & 6.6$^{+0.5}_{-0.3}\times$ 10$^{17}$ & 1.3$^{+0.1}_{-0.2}\times$ 10$^{17}$ & 326$^{+8}_{-6}$ & 263$^{+14}_{-9}$ \\
C$_2$H$_5$OH;v=0; & 4.3$^{+0.4}_{-0.4}\times$ 10$^{17}$ & 1.2$^{+0.1}_{-0.2}\times$ 10$^{17}$ & 335$^{+12}_{-12}$ & 306$^{+12}_{-14}$ \\
C$_2$H$_5$CN;v=0; & 1.7$^{+0.1}_{-0.1}\times$ 10$^{16}$ & \nodata & 320$^{+10}_{-8}$ & \nodata \\
CH$_3$COCH$_3$;v=0; & 7.7$^{+1.3}_{-0.8}\times$ 10$^{17}$ & \nodata & 305$^{+6}_{-14}$ & \nodata
\enddata
\tablecomments{$\dagger$The temperature is assumed to be 200~K in the simulations. }
\end{deluxetable}

\begin{figure*}[htb!]
\centering
\includegraphics[width=\linewidth]{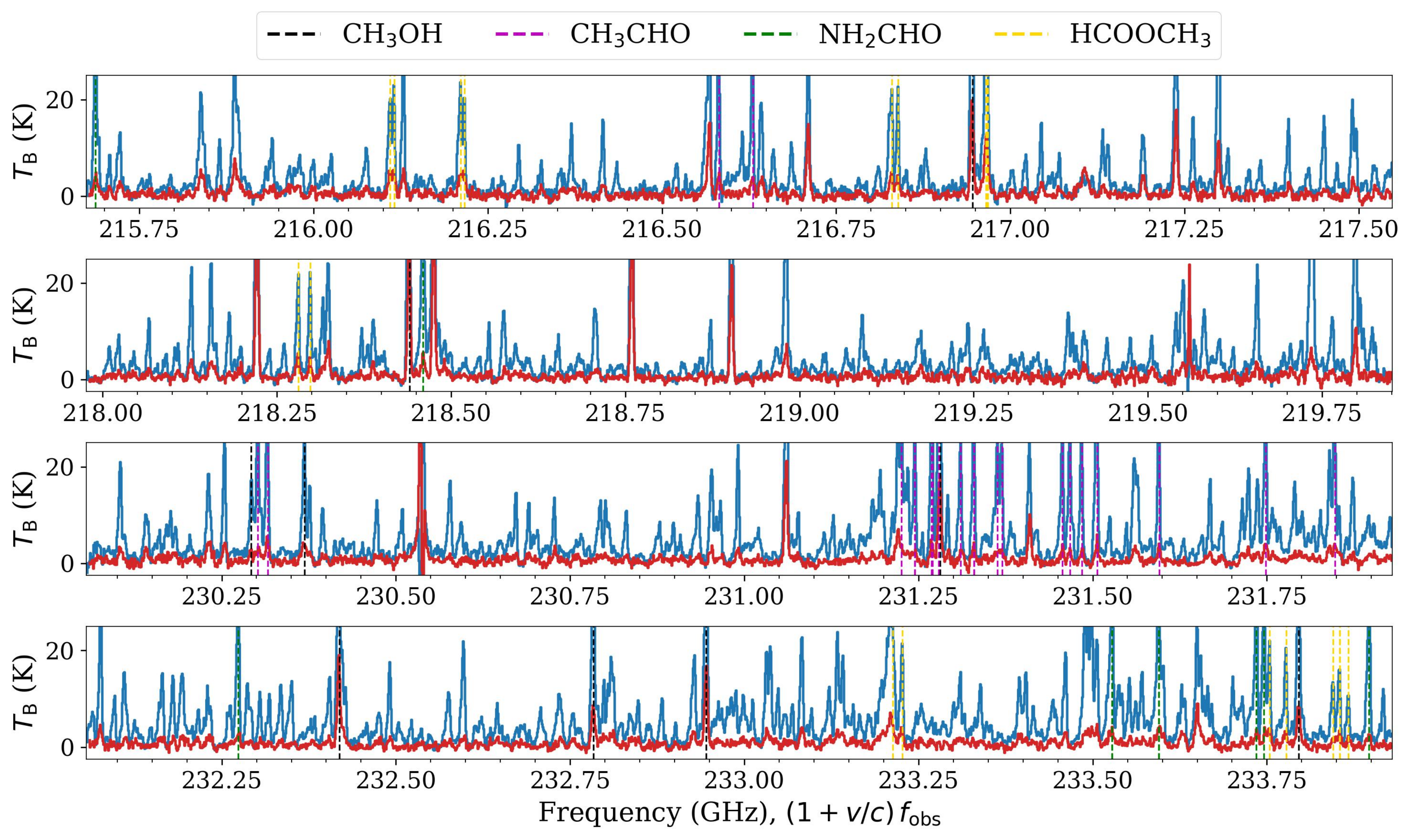}
\caption{\label{fig:specK_selected} 
The spectra toward the continuum peaks of source A (blue) and B (red). 
The dashed lines illustrate the selected transitions of COMs. 
The x axis is the frequency shifted back to the expected rest frequency with $v=2.0$ and $6.5$ km s$^{-1}$ for source A and source B, respectively.  
}
\end{figure*}

\section{Chemical Compositions of the Hot Corinos}
\label{sec:chem}

\subsection{Molecular Spectrum Synthesization}
\label{sec:chem:methods}

We further examined the molecular compositions, particularly the COMs, in the two sources using the wide-bandwidth data from project \#2018.1.00302.S. 
At the angular resolution of these data (0\farcs37), the COM emission shows compact, core-like morphologies, with slight variations in peak positions among species. 
To minimize mutual contamination between the two sources, given the comparable beam size (0\farcs37) and their close separation (0\farcs5), we extracted spectra at the dust continuum peaks of each source.
As shown in Figure~\ref{fig:specK_selected}, the two spectra shows a clear velocity offset, justifying that the emission from the two sources are well-separated. 
Both spectra show a forest of lines, while source A is apparently line-richer than source B. 
As labeled in the figure, lots of lines are the transitions of COMs, including CH$_3$OH, CH$_3$CHO, HCOOCH$_3$, and NH$_2$CHO, indicating that both source A and source B are rich in complex organic chemistry.
Since line intensity depends on the gas density in each source, we need to derive the relative column densities in order to compare the COM abundances between the two sources.

We used \texttt{XCLASS} \citep[eXtended CASA Line Analysis Software Suite;][]{2017Moller_XCLASS} to evaluate the excitation temperature and total column density for each molecular species.
\texttt{XCLASS} calculates the synthetic spectra of molecular species under the assumption of local thermodynamic equilibrium (LTE), given molecular parameter inputs such as source size, excitation temperature (\Tex), total column density (\Ntot), line width (\vwidth), and velocity offset (\voff).
To reduce degeneracy in the parameter fitting, we fixed the source size to 0\farcs37, the line width FWHM to 5 km s$^{-1}$, and the velocity offsets to 2.0 and 6.5 km s$^{-1}$ for source A and source B, respectively.
Both the line width FWHM and the velocity offsets were estimated based on the bright CH$_3$OH lines, which were further examined by the synthesized spectra. 
For reference, the wide-bandwidth data (\#2018.1.00302.S) have a spectral resolution of 1.4 km s$^{-1}$.
The derivation of excitation temperature (\Tex) and total column density (\Ntot) was performed in three steps.
\begin{enumerate}
    \item Initial Estimation. 
    We fixed the \Tex\ to 200~K and manually adjusted the \Ntot\ to obtain a rough match with the observed line strengths.
    \item Independent Fitting per Species. 
    For each molecular species, we independently optimized \Tex\ and \Ntot\ using the ``Nested Sampling (NS)'' algorithm. 
    The \Tex\ was allowed to vary between 100 and 300 K. 
    The \Ntot\ was allowed to vary between 0.01 and 100 times the initial value obtained in Step 1.
    \item Simultaneous Multi-Species Fitting. 
    All molecular species were then fitted simultaneously using the NS algorithm. 
    The \Tex\ was allowed to vary within $\pm$50 K of the values derived in Step 2. 
    The \Ntot\ were allowed to vary between 0.1 and 10 times the corresponding Step 2 values.
    \item Error Estimation. 
    Finally, we estimated the uncertainties of the values derived from Step3 using the ``Markov chain Monte Carlo (MCMC)'' method. 
\end{enumerate}
Table~\ref{tab:molec} summarizes the resulting excitation temperature \Tex\ and total column density \Ntot\ of each molecular species in the two sources. 
Figures~\ref{fig:appx_spec_spw3}-\ref{fig:appx_spec_spw1} in appendix show the observed and the synthesized spectra for both sources. 
Table~\ref{tab:appx:trans:almasop} in appendix lists the transition parameters. 
We note that in source A, the COM emission extends slightly beyond the beam, meaning that our analysis captures only the bright central region. 
As a result, the derived column densities for source A are possibly underestimated relative to the full spatial extent of its COM emission.
Particularly, CH$_3$OH extends beyond the beam, exhibits optical thickness in low-$E_u$ transitions, and is additionally traced by an intervening component. 
Furthermore, in the future a more precise estimation could be achieved using a customized source size for each molecule, as the molecule extents are not uniform. 

Among the COMs, we note that CH$_3$OH transitions with low upper-state energies are likely optically thick, as suggested by the images (Figure \ref{fig:mom0_CH3OH}).
In addition, high–upper-energy transitions of CH$_3$OH may be affected by radiative pumping in addition to collisional excitation.
These effects introduce additional uncertainty into the derivation of CH$_3$OH column densities.

\subsection{Column Densities of COMs}
\label{sec:chem:colDens}

The $^{12}$C/$^{13}$C ratios of CH$_3$OH, derived from the column density ratio between CH$_3$OH $v=0$ and $^{13}$CH$_3$OH, are 8 and 13 for source A and source B, respectively. 
These ratios are significantly lower than the local ISM value of $\sim$50--70 and are often interpreted as evidence that the CH$_3$OH $v=0$ emission is optically thick \citep[e.g., ][]{2013Zapata_IRAS16293-2322B,2019Lee_HH212_COM_atm,2022Hsu_ALMASOP}. 
The D/H ratios of CH$_3$OH, derived from the column density ratio between CH$_2$DOH and CH$_3$OH $v=0$ are 0.4 and 0.1 for source A and source B, respectively. 
Adopting the column density of $^{13}$CH$_3$OH with a $^{12}$C/$^{13}$C ratio of 68 \citep{2005Milam_12C13C}, the D/H ratios become 0.05 and 0.02 for source A and source B, respectively. 
We note that it remain uncertain whether the low $^{12}$C/$^{13}$C ratios found in hot corinos result (solely) from optically thick CH$_3$OH $v=0$ emission. 
See Sect.~\ref{sec:disc:12C13C} for more discussion. 

\begin{figure}[htb!]
\centering
\includegraphics[width=.9\linewidth]{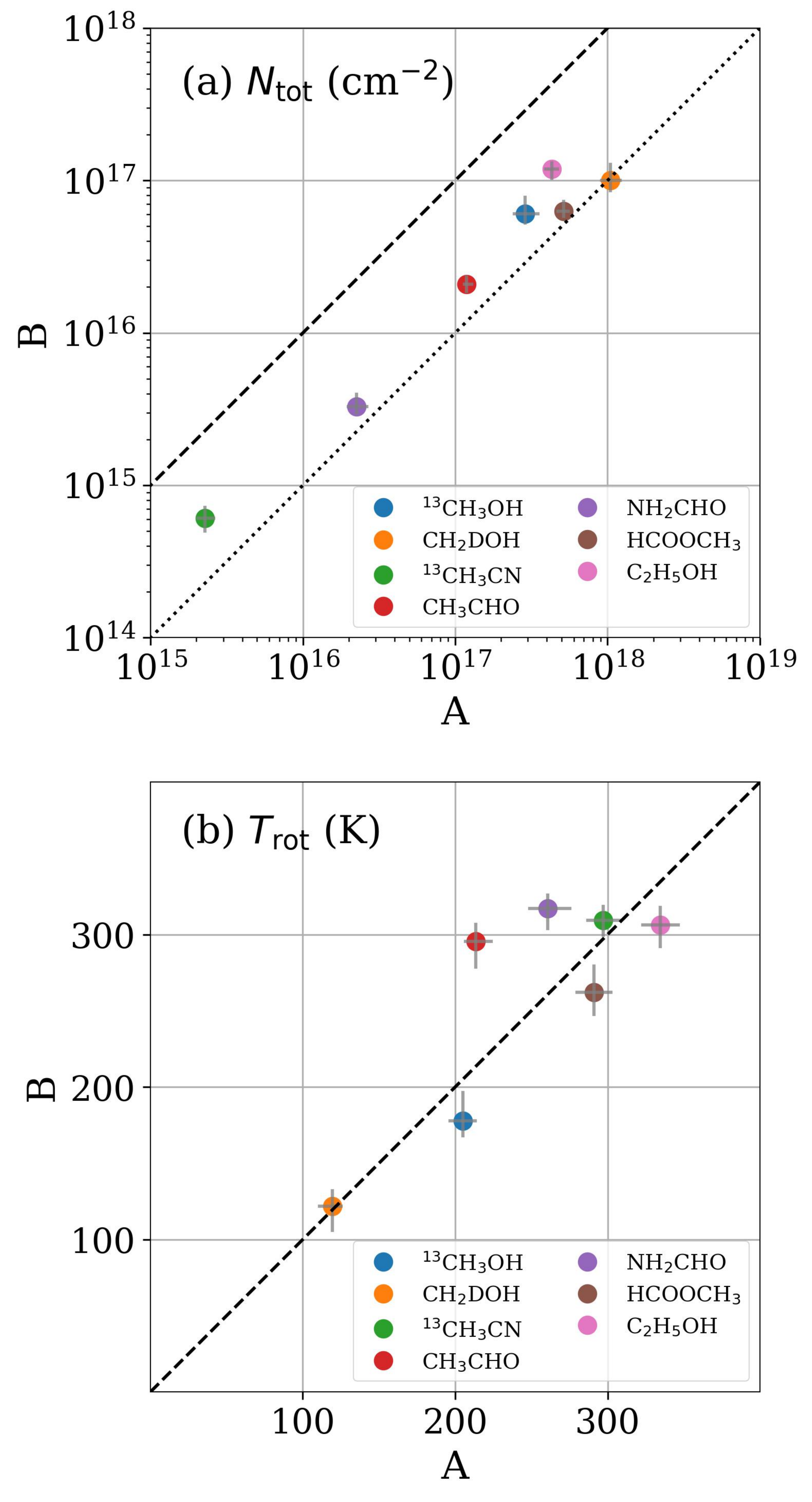}
\caption{\label{fig:colDens_all} 
The column densities (a) and rotational temperature (b) distributions of selected COMs in source A and source B. 
In panel (a), the dashed and dotted lines illustrate the relations of $y=x$ and $y=0.1x$, respectively. 
In panel (b), the dashed line illustrate the relations of $y=x$. 
}
\end{figure}

In Figure~\ref{fig:colDens_all}(a), we show the column density distributions of COMs for source A and source B based on their ground vibration state transitions. 
Each marker represents a COM species and the two lines indicate the proportional correlations with factors of 1 and 0.1. 
Note that the main isotopologue of methanol, CH$_3$OH, is excluded, due to significant contributions from the intervening component. 
As shown in Figure~\ref{fig:colDens_all}(a), the column densities of source A and source B exhibit a linear pattern, indicating a comparable COM column density ratio $\sim0.1$ between the two sources.  
This fixed ratio suggests that the COM composition in source A and source B hot corinos is similar. 
The higher COM column densities in source A can be attributed to its higher gas density relative to source B.
For a given fractional abundance, the total number of COM molecules scales positively with gas density.
Consistently, all COM lines detected in source B are also detected in source A.
The COM lines detected only in source A may simply remain undetected in source B due to the lower SNR.
The only line that is stronger in source B than source A is SiO ($\sim$217.1 GHz in Figure~\ref{fig:specK_selected}), which is not a COM. 

\subsection{Rotational Temperatures of COMs }
\label{sec:chem:temperature}

In Figure~\ref{fig:colDens_all}(b), we show the rotational temperature distributions of COMs for source A and source B based on their ground vibration state transitions. 
The COMs have warm temperatures of around 100--350~K, consistent with the characteristic of a hot corino. 
In general, the rotational temperatures in source A are comparable to those in source B, suggesting that the COMs in the two hot corinos reside in similar thermal conditions. 
Meanwhile, the distributions can be tentatively divided into three groups. 
CH$_2$DOH has the lowest temperature of $\sim$100~K, $^{13}$CH$_3$OH and CH$_3$CHO have the intermediate temperature of $\sim$200~K, and the others have the highest temperature of $\sim$300~K. 
This thermally onion-like distribution could potentially infer their different extents. 
For example, NH$_2$CHO was reported to be is found to reside in more compact regions \citep{2020Manigand_IRAS16293_COM,2022Bianchi_SVS13A_COM,2022Lee_HH212_stratification,2022Okoda_B335_chem,2025Frediani_IRAS4A2_chem}. 


\section{Discussion}
\label{sec:Discussions}

\subsection{Origins of the COM-rich Features in HOPS-288}
\label{sec:disc:origin}

Our findings suggest three COM-rich features in HOPS-288: a localized warm region in source A, a localized warm region in source B, and a feature between source A and source B (intervening CH$_3$OH-rich feature). 

\subsubsection{Hot Corino in source A}

The COM emission from source A likely traces the hot inner disk, though higher resolution observations are still necessary for further confirmation. 
Such COM-rich features associated with disks in Class 0/I protostellar cores are rarely reported, with only a few known cases such as HH212 \citep[Class 0;][]{2016Codella_HH212_H2O_COM,2017Lee_HH212,2019Lee_HH212_COM_atm,2022Lee_HH212_stratification} and V883 Ori \citep[Class I/II;][]{2018vantHoff_V883Ori_CH3OH,2019Lee_V883-Ori_outburst,2025Jeong_V883Ori_COM,2025Fadul_V883Ori}.
Detecting COM gas in disks is challenging because the warm region is extremely compact, and the dusty disk blocks irradiation from the central protostar, preventing most of the outer disk from heating \citep{2022Nazari_model_diskshadow,2023Hsu_ALMASOP}.
For HH 212, the warm disk has been attributed to direct heating from the central protostar or to interactions with disk winds \citep{2022Lee_HH212_stratification}.
For V883 Ori, the presence of a warm inner disk is thought to result from a stellar outburst \citep{2018vantHoff_V883Ori_CH3OH,2019Lee_V883-Ori_outburst,2025Jeong_V883Ori_COM}.

Given that the modeled stellar mass of source A suggests it is an intermediate-mass protostar as well as the observed high bolometric luminosity \citep[$180\pm70$\Lsun,][]{2020Dutta_ALMASOP,2022Hsu_ALMASOP}, we expect that the irradiation from the central protostar predominantly heats the disk.
Meanwhile, the possible additional heating source includes the streamer shock from the circumbinary disk to the circumstellar disk \citep{2014Tang_UYAurigae} and viscous dissipation within the circumstellar disk \citep{2021Liu_FuOriDisk}. 
Furthermore, the binarity of source A itself could possibly bring more complication, as the COM emission can possibly be originated from a circumbinary disk. 

On the other hand, the binarity of source A may also be relevant to the asymmetric morphology of COM emission in source A (Figure~\ref{fig:mom0_CH3OH} and Figure~\ref{fig:mom0_COM}). 
In source A, as shown in Figure~48 of \citet{2020Tobin_VANDAM_disk}, the northwestern substructure is the primary component in the source A, consistent with the asymmetric COM morphology of this study. 

\subsubsection{Hot corino in source B}

Due to the limited spatial resolution, we are not able to resolve the structure of the hot corino in source B, while its rotation motion is suggested by a tentative velocity gradient.
Such rotating feature can possibly be an unresolved disk or a ring \citep[e.g., ][]{2025Hsu_G192.12-11.10_COM}. 
The rotating ring can be attributed to the accretion shocks when materials enter the disk. 
Observations of COMs in other systems seem to trace the accretion shocks include L1157 \cite{2002Velusamy_L1157}, IRAS 16293–2422 A \citet{2016Oya_IRAS-16293-2422}, B335 \citep{2022Okoda_B335_chem}, and G192.12-11.10 \citep{2025Hsu_G192.12-11.10_COM}. 
On another note, the shock heating on the surface of a pseudodisk is also speculated to release COMs \citep{2025Wang_pseudodisk}. 

\subsubsection{CH$_3$OH-rich feature between source A and source B}

In addition to the two hot corinos, we identify a CH$_3$OH-rich feature, most prominently traced in low-excitation transitions and with tentative CH$_3$CHO emission.
This intervening component is located between source A and source B, appearing to connect source A to source B, with its peak closer to the latter.
The structure may trace a shocked region within the circumbinary disk. 
Alternatively, it could represent a bridge linking the two sources, if the observed tail-like morphology is not a consequence of beam convolution. 

The association of CH$_3$OH with streamer-induced shocks onto circumbinary disks has been reported in the proto-binary system [BHB2007] 11, where \citet{2022Vastel_BHB2007-11_COMs} identified three distinct CH$_3$OH components with different positions and velocities. 
They proposed that one arises from a localized region around one source, while the other two reside in the circumbinary disk and are associated with shocks driven by mass-accretion streamers. 
However, they also noted that alternative explanations could not be excluded, partly due to the limited spatial resolution.
In our target, both the morphology and kinematics of this tail are clearly observed. 
The velocities and spatial extent of CH$_3$OH are consistent with the inner redshifted portion of the circumbinary disk (Figure~\ref{fig:PV_tail}). 

A bridge has been observed in the proto-binary system IRAS 16293–2422 between the two members having a separation of 620~au \citep[e.g.,][]{2012Pineda_IRAS16293_bridge, 2019vanderWiel_IRAS16293_bridge}.
In that source, the bridge is traced by dust continuum and relatively simple molecules such as C$^{17}$O, C$^{18}$O, H$_2$CO, and H$^{13}$CN \citep{2012Pineda_IRAS16293_bridge,2019vanderWiel_IRAS16293_bridge}, whereas COMs, including CH$_3$OH, appear only at the two continuum peaks rather than along the bridge \citep[e.g.,][]{2016Jorgensen_PILS}.
Furthermore, the bridge was found to be kinematically quiescent \citep{2019vanderWiel_IRAS16293_bridge}.
In contrast, in our target HOPS-288, the tail is not detected in dust continuum but is instead traced by CH$_3$OH. 
Also, the separation is around 200~au, much closer than that in IRAS 16293–2422. 
It is uncertain, however, as to how a quiescent bridge could lead to the desorption of CH$_3$OH. 
Therefore, the quiescent bridge scenario seen in IRAS 16293–2422 does not match HOPS-288. 
Regarding the scenario in which CH$_3$OH in HOPS-288 traces a dynamically active bridge, the enhancement of CH$_3$OH closer to source B could indicate a shock where material is flowing from source A toward source B. 
Given that source A is more massive than source B (Sect.~\ref{sec:CH3OH:PV}), this would imply that the primary (more-massive) member is ``feeding'' the secondary (less-massive) one. 


\subsection{Chemical Similarities among Proto-binary/multiple Systems}

\subsubsection{Hot corinos exhibiting rotation motion in the Orion}

\begin{figure}[htb!]
\centering
\includegraphics[width=\linewidth]{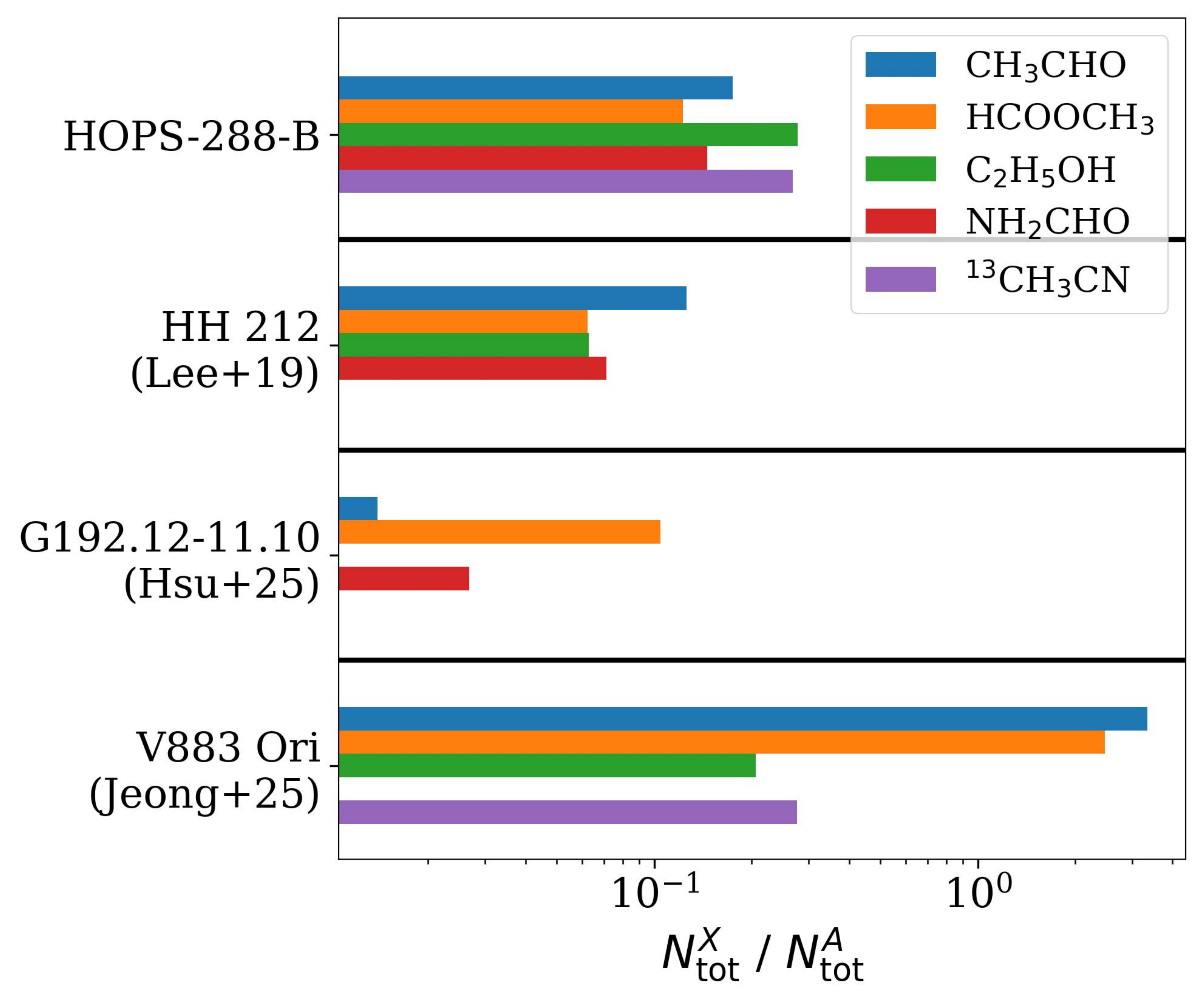}
\caption{\label{fig:colDensRatio_lit} 
Column density ratios of different disk-origin hot corinos in Class 0/I protostellar cores and HOPS-288-A (source A). 
The protostellar core names are labeled in y-axis. 
Each color represents a molecular species. 
Comparing the relative ratios within each group provides useful insights into the chemical similarities between the labeled sourced and HOPS-288-A (source A). 
}
\end{figure}

As shown in Figure~\ref{fig:colDens_all} and described in Sect.~\ref{sec:chem:colDens}, source A and source B appear to have similar chemical compositions. 
In Figure~\ref{fig:colDensRatio_lit}, we present the column density ratios of source B relative to source A for the commonly detected COMs.
Each bar represents the column density ratio for an individual species.
As shown in Figure~\ref{fig:colDensRatio_lit}, these ratios cluster around a value of $\sim$0.1--0.2.
We note that the column densities were derived assuming a fixed molecular source size for both source A and source B.
Consequently, any difference in the actual spatial extents of the two hot corinos would introduce an additional factor (i.e., the ratio of their beam dilution) into the derived column density ratios.
For example, as source A is expected to have a larger hot corino than source B, the column density ratios could be underestimated. 

To further examine the observed chemical similarity between source A and source B, we compare the COM column densities between source A and those of other hot corinos in Orion that are known to exhibit rotation motions or be disk-origin: HH~212 \citep{2017Lee_HH212,2019Lee_HH212_COM_atm}, G192.12-11.10 \citep{2022Hsu_ALMASOP,2025Hsu_G192.12-11.10_COM}, and V883~Ori \citep{2018vantHoff_V883Ori_CH3OH,2019Lee_V883-Ori_outburst,2025Jeong_V883Ori_COM,2025Fadul_V883Ori}.
We adopt column densities from \citet{2019Lee_HH212_COM_atm} for HH~212, \citet{2022Hsu_ALMASOP} for G192.12-11.10, and \citet{2025Jeong_V883Ori_COM} for V883~Ori.
Comparing the relative column density ratios between different sources provides useful insights on chemical similarity, though variations in observed transitions and analysis methodologies introduce some uncertainty.
The methods used to derive the column densities in each study are summarized in Table~\ref{tab:lit}.
In \citet{2022Hsu_ALMASOP}, spectra were extracted at the dust continuum peak using a beam that encloses the COM emission, while in the other studies, extraction regions were chosen to adequately encompass the emitting areas (for HH~212, only the lower disk atmosphere was used).

As shown in Figure~\ref{fig:colDensRatio_lit}, in the disk of the Class 0 protostellar core HH 212, CH$_3$CHO is significantly more abundant than the other O-bearing (HCOOCH$_3$) and N-bearing (NH$_2$CHO) COMs.
In the disk of the Class I/II protostellar core V883 Ori, CH$_3$CHO and HCOOCH$_3$ are enhanced relative to C$_2$H$_5$OH, and $^{13}$CH$_3$CN. 
In contrast, the column density ratios of COMs between source B and source A are very close (around 0.2),  supporting the chemical similarities among proto-binary or multiple systems.

\begin{deluxetable*}{llllllll}
\tablefontsize{\normalsize}
\tablecaption{\label{tab:lit} The information of the studies of hot corinos in the Orion cloud and exhibiting rotation motion or disk-originated. }
\tablehead{
\colhead{Reference} & \colhead{Source} & \colhead{Method} & \colhead{Fixed Parameters} 
}
\startdata
This study                    & HOPS-288      & Spectrum Fitting   & $\delta v=5$ km s$^{-1}$ \\ 
\citet{2019Lee_HH212_COM_atm} & HH 212        & Rotational Diagram (Optically thin) & \Trot$=150$~K \\ 
\citet{2022Hsu_ALMASOP}       & G192.12-11.10 & Spectrum Fitting   & \\ 
\citet{2025Jeong_V883Ori_COM} & V883 Ori      & Spectrum Fitting   & \Trot$=120$~K; $\delta v=3$ km s$^{-1}$ \\ 
\enddata
\end{deluxetable*}

\subsubsection{Chemical Similarities and Diversities among Proto-binary/multiple Systems}

To our knowledge, such detailed comparisons of complex organics in proto-binary or multiple systems remain rare. 
In several systems, including L1551 IRS5 \citep[50 au;][]{2020Bianchi_L1551-IRS5_COM}, L1448 IRS3B \citep[200 au;][]{2021Yang_PEACHES}, Ser-emb 11 \citep[888 au;][]{2021Martin_Ser-emb-11_COM}, and [BHB2007] 11 \citep{2022Vastel_BHB2007-11_COMs}, hot corinos are detected in only one member. 
Non-detections typically result from low envelope gas densities or low luminosities, and may also be influenced by factors such as dust opacity \citep{2019Sahu_IRAS4A_hot-corino-atm,2023Hsu_ALMASOP}.
In NGC 1333 IRAS 4A (470 au), both members host COMs, but measurements toward one member are challenging due to dust opacity \citep{2017Lopez_IRAS4A_COM,2019Sahu_IRAS4A_hot-corino-atm,2020DeSimone_IRAS4A_COM}. 
In IRAS 16293–2422, \citet{2020Manigand_IRAS16293_COM} compared column density ratios between sources A and B and did not find overall similarities. 
Given the relatively wide separation of $\sim$620~au, it remains unclear how chemically similar such systems can inherit from their parental cloud.

One suitable comparison object for HOPS-288 is SVS 13A, where the two members are separated by only $\sim$90~au and COMs have been detected toward both VLA4A and VLA4B \citep[e.g.,][]{2019Bianchi_SVS13A_COM,2021Yang_PEACHES,2022Bianchi_SVS13A_COM,2025Hsieh_SVS13A_12C13C}. 
\citet{2022Bianchi_SVS13A_COM} reported COM abundances for VLA4A and VLA4B individually, adopting fixed rotational temperatures for each member and evaluating column densities for CH$_3$CHO (two lines), NH$_2$CHO (three lines), and CH$_3$OCH$_3$ (16 lines). 
The resulting column density ratios of CH$_3$CHO, NH$_2$CHO, and CH$_3$OCH$_3$ between the two members were 0.46, 0.24, and 1.33, respectively. 
As pointed out by \citet{2022Bianchi_SVS13A_COM}, their observed chemical segregation between the two members may result from the onion-like structure of the two hot corinos.
For example, compared to the other COMs, NH$_2$CHO is formed in a compact region and consequently be more easily obscured by the dust. 
Therefore, its column density in the member having more optically thick dust continuum (VLA4B) could be more underestimated than the other one (VLA4A). 
Given the limited number of proto-binary/multiple systems studied in detail, the question of how chemical compositions compare between members within proto-binary/multiple systems remains largely open.

\subsubsection{Implications}

Here we propose three scenarios to interpret the chemical similarities in COMs between the YSOs within the close binary/multiple system HOPS-288: (i) early chemical imprint, (ii) common reservoir accretion, and (iii) inter-source exchange.
First, 
if source A and source B are evolving independently, the observed similarity would suggest that the chemical composition is established even earlier, possibly during the starless core phase. 
Considering that source A may be at a more evolved stage than source B (Section~\ref{sec:outflow}), the observed chemical similarity between them may further indicate that the chemistry of commonly detected COMs does not change substantially with protostellar evolution at these early stages.
Second, 
the similarities may reflect a common chemical reservoir, where material is funneled toward the two hot corinos. 
The reservoir could be either the circumbinary disk or gas components at an even larger scale. 
The role of streamers in shaping COM chemistry has been suggested, for example, to explain the unusual $^{12}$C/$^{13}$C ratio of methyl cyanide in SVS 13A \citep{2025Hsieh_SVS13A_12C13C}, despite the other scenarios were also proposed \citep{2025Busch_SVS13A_12C13C}. 
If streamers can indeed significantly influence the COM composition of the accreting protostellar core, then the observed COM similarity would imply that the streamers themselves share chemically similar, or even identical, origins.
Finally, 
the similarities may arise from material exchange between the two members. 
In this scenario, comparable chemical compositions would be naturally expected, as one member accretes material from the other.
The intervening component identified in this study between source A and source B could potentially trace such interactions, serving as a bridge for the transfer of material between the two sources.

The three proposed scenarios are not mutually exclusive. 
Namely, the two sources could inherit similar materials from their parent cloud, absorb the materials from a coherent nearby reservoir, and then exchange material with each other. 


\subsubsection{Caveats}
The reported chemical similarities are based on the derived column densities, which could be affected by several factors. 

Although we often refer to the warm region rich in COMs surrounding a solar-type protostar as a hot corino, the spatial distributions of individual COMs within this region may differ \citep[e.g.,][]{2022Bianchi_SVS13A_COM,2024Hsu_MMS6}.
In reality, different COM species can have distinct emitting extents, line widths, and even velocity centroids.
However, in our analysis, these parameters were fixed, which could introduce uncertainties in the derived column densities.
For example, a component size of $\sim$0\farcs2 (the deconvolved continuum size of source B) could possibly lead to an underestimation of column densities by a factor of two. 
Meanwhile, we do not account for the potential effect of dust opacity, which could lead to underestimated molecular column densities.
A well-known example is NGC 1333 IRAS 4A1, where \citet{2019Sahu_IRAS4A_hot-corino-atm} detected COM absorption lines at millimeter wavelengths and proposed that dust opacity was responsible.
Subsequently, \citet{2020DeSimone_IRAS4A_COM} revealed COM emission from the same source at centimeter wavelengths, further supporting the influence of dust opacity. 
Our target, HOPS-288, is likely a nearly edge-on system and therefore could also be affected by dust opacity.
A similar situation is seen in the nearly edge-on disk of HH 212, where COM emission exhibits a ``hamburger-like'' morphology due to the midplane being thermalized by optically thick dust \citep{2017Lee_HH212,2019Lee_HH212_COM_atm,2022Lee_HH212_stratification}. 
Detecting gas-phase COMs in disks is additionally difficult due to the small warm region \citep[e.g., ][]{2022Nazari_model_diskshadow,2023Hsu_ALMASOP}. 
As a result, the detected COMs associated with disks are often at the periphery of the disk, such as the surface atmosphere in HH 212 and the disk-envelope interface \citep[e.g., ][]{2016Oya_IRAS-16293-2422,2022Okoda_B335_chem,2025Hsu_G192.12-11.10_COM}. 
Similarly, \citet{2022Bianchi_SVS13A_COM} suggested that the observed chemical segregation in SVS 13A may be linked to the layered spatial extents of COMs, as well as to dust opacity effects.
In our observations, the hot corinos are not spatially well resolved and we cannot rule out the role of the dust opacity. 

\subsection{$^{12}$C/$^{13}$C ratio of CH$_3$OH}
\label{sec:disc:12C13C}

As shown in Sect.~\ref{sec:chem:colDens}, the $^{12}$C/$^{13}$C ratios of CH$_3$OH are 12 and 8 for source A and source B, respectively, indicating that CH$_3$OH $v=0$ emission is optically thick. 
However, similarly low $^{12}$C/$^{13}$C ratios of CH$_3$OH are commonly observed in hot corinos \citep[e.g., ][]{2022Hsu_ALMASOP,2025Busch_SVS13A_12C13C}. 
Moreover, \citep{2023Lee_HOPS373SW} reported that even ratios derived from optically thin transitions of CH$_3$OH $v=2$ remain as low as about 20.
This raises the question of whether the generally low $^{12}$C/$^{13}$C ratios result (solely) from optically thick CH$_3$OH $v=0$ emission. 
To address this, \citet{2025Busch_SVS13A_12C13C} suggests that $^{13}$C enrichment in COMs during early protostellar stages can be inherited from precursor species. 
The $^{12}$C/$^{13}$C ratios in the prestellar phase are established through isotopic exchange reactions.
More studies are necessary for further confirmations. 


\subsection{Implications of Formamide}
\label{sec:disc:formamide}

NH$_2$CHO is suggested as a pre-biotic precursor \citep[e.g., ][]{2012Saladino_formamide,2019Lopez_formamide}. 
In Figures~\ref{fig:mom0_COM} and \ref{fig:PV_COM_Cmajor}, NH$_2$CHO is found to reside in much more compact regions than the other COMs. 
Similar behavior has been reported in, for example, IRAS 16293-2422 A \citep{2020Manigand_IRAS16293_COM} and HH 212 \citep{2022Lee_HH212_stratification}. 
\citet{2022Lee_HH212_stratification} showed that the extents of NH$_2$CHO, CH$_3$OH, and H$_2$CO are anti-correlated with their binding energies, consistent with these molecules being thermally desorbed from icy mantles in the disk. 
Beyond the thermal desorption of NH$_2$CHO from dust-grain mantles, its compact morphology in HOPS-288 may also suggest that NH$_2$CHO formation on ices is enhanced in the warmer, inner regions. 
This could be related to external vacuum-ultraviolet (VUV) irradiation, as laboratory experiments by \citet{2020Martin-Domenech_NH2CHO} demonstrated that NH$_2$CHO formation from radicals can be significantly boosted under VUV irradiation. 
On the other hand, \citet{2024Lopez-Sepulcre_L1157_formamide} demonstrated that the NH$_2$CHO observed in the outflow L1157-B can be largely explained by gas-phase chemistry, specifically via the reaction H$_2$CO + NH$_2$ $\rightarrow$ NH$_2$CHO + H$_2$. 
However, since their modeling was tailored to the outflow environment (e.g., molecular hydrogen number density of $4\times10^5$ cm$^{-3}$), the same conclusion may not be directly applicable to disk environments such as the hot corinos in HH 212 and HOPS-288. 


\section{Conclusions}
\label{sec:Conclusions}

Utilizing two ALMA programs, \#2018.1.01038.S and \#2018.1.00302.S, we investigated the physical structure as well as the morphology, kinematics, and compositions of COMs in the hierarchical 2+1 proto-triple system HOPS-288.
The system was treated as a proto-binary, consisting of HOPS-288-A and HOPS-288-B, since the local binarity of the former can not be resolved.

\begin{enumerate}  
\item 
We identified two CO outflows, each associated with one source, along with a C$^{18}$O circumbinary disk. 

\item 
We revealed three COM-rich features in HOPS-288: two hot corinos associated with HOPS-288-A and HOPS-288-B, and an intervening component between them.
The intervening component is traced by CH$_3$OH (particularly its low-$E_u$ transitions) and tentatively by CH$_3$CHO, whereas the other COMs appear confined to the two hot corinos.
Both the hot corinos in HOPS-288-A and HOPS-288-B are likely tracing rotation motions. 
The kinematics of the former further suggests its disk origin.  
The latter is indicated by a tentative velocity gradient along its major axis. 
An intervening component may trace a shocked region within the circumbinary disk induced by a mass accretion streamer. 
Alternatively, if the tail-like morphology is not resulting from beam convolution, this component could primarily represent a bridge linking the two sources. 

\item 
We derived the rotational temperatures and column densities of molecules in the inner part of HOPS-288-A and HOPS-288-B. 
The derived column density ratios of COMs are similar between the two sources, suggesting comparable complex organic chemistry. 
Additional uncertainties could be possibly caused by dust opacity and the diverse COM extents.
We propose three possible scenarios to explain these similarities: (i) inter-source exchange, (ii) common reservoir accretion, and (iii) early chemical imprint. 

\item 
The $^{12}$C/$^{13}$C ratios of CH$_3$OH, derived from the column density ratios of CH$_3$OH $v=0$ to $^{13}$CH$_3$OH, are $\sim$10, significantly below the interstellar medium value of $\sim$50--70.
The D/H ratios of CH$_3$OH, calculated from the column density ratios of CH$_2$DOH to CH$_3$OH $v=0$, are 0.4 and 0.1 for sources A and B, respectively.
Assuming a $^{12}$C/$^{13}$C ratio of 68 and using the $^{13}$CH$_3$OH column density, the adjusted D/H ratios are 0.05 and 0.02 for sources A and B, respectively.
Low $^{12}$C/$^{13}$C ratios in CH$_3$OH are commonly observed in hot corinos, but it remains unclear whether these low values result solely from optically thick emission in CH$_3$OH $v=0$.

\item 
Within the hot corino in HOPS-288-A, the COM emission shows an asymmetric brightness distribution: the blueshifted side is systematically brighter than the redshifted side.
Such effect is less pronounced in HCOOCH$_3$. 
In addition, NH$_2$CHO emission is concentrated in more compact, inner regions compared to other COMs

\end{enumerate}


\acknowledgments
This paper makes use of the following ALMA data: ADS/JAO.ALMA\#2018.1.00302.S and ADS/JAO.ALMA\#2018.1.01038.S. ALMA is a partnership of ESO (representing its member states), NSF (USA) and NINS (Japan), together with NRC (Canada), NSTC and ASIAA (Taiwan), and KASI (Republic of Korea), in cooperation with the Republic of Chile. The Joint ALMA Observatory is operated by ESO, AUI/NRAO and NAOJ.
This work made use of Astropy:\footnote{http://www.astropy.org} a community-developed core Python package and an ecosystem of tools and resources for astronomy \citep{astropy:2013, astropy:2018, astropy:2022}. 
S.-Y.H. acknowledges support from the Academia Sinica of Taiwan (grant No. AS-PD-1142-M02-2).
S.-Y. H. and C.-F.L. acknowledge grant from the National Science and Technology Council of Taiwan (112-2112-M-001- 039-MY3).  
N.M.M. acknowledges support from the DGAPA–PAPIIT IA103025 grant. 
D.J.\ is supported by NRC Canada and by an NSERC Discovery Grant. 
Supported by the international Gemini Observatory, a program of NSF’s NOIRLab, which is managed by the Association of Universities for Research in Astronomy (AURA) under a cooperative agreement with the National Science Foundation, on behalf of the Gemini partnership of Argentina, Brazil, Canada, Chile, the Republic of Korea, and the United States of America.
L. B. gratefully acknowledges support by the ANID BASAL project FB210003. 
QY-L acknowledges the support by JSPS KAKENHI Grant Number JP23K20035.

\software{
Astropy \citep{astropy:2013, astropy:2018, astropy:2022},
\texttt{CASA} \citep{casa:2022},
\texttt{CARTA}  \citep{2021Comrie_CARTA}, 
\texttt{XCLASS} \citep{2017Moller_XCLASS}.
}

\clearpage
\appendix
\restartappendixnumbering

\section{Molecular Transitions \label{appx:trans}}
\resetapptablenumbers

\begin{figure*}[htb!]
\centering
\includegraphics[width=\linewidth]{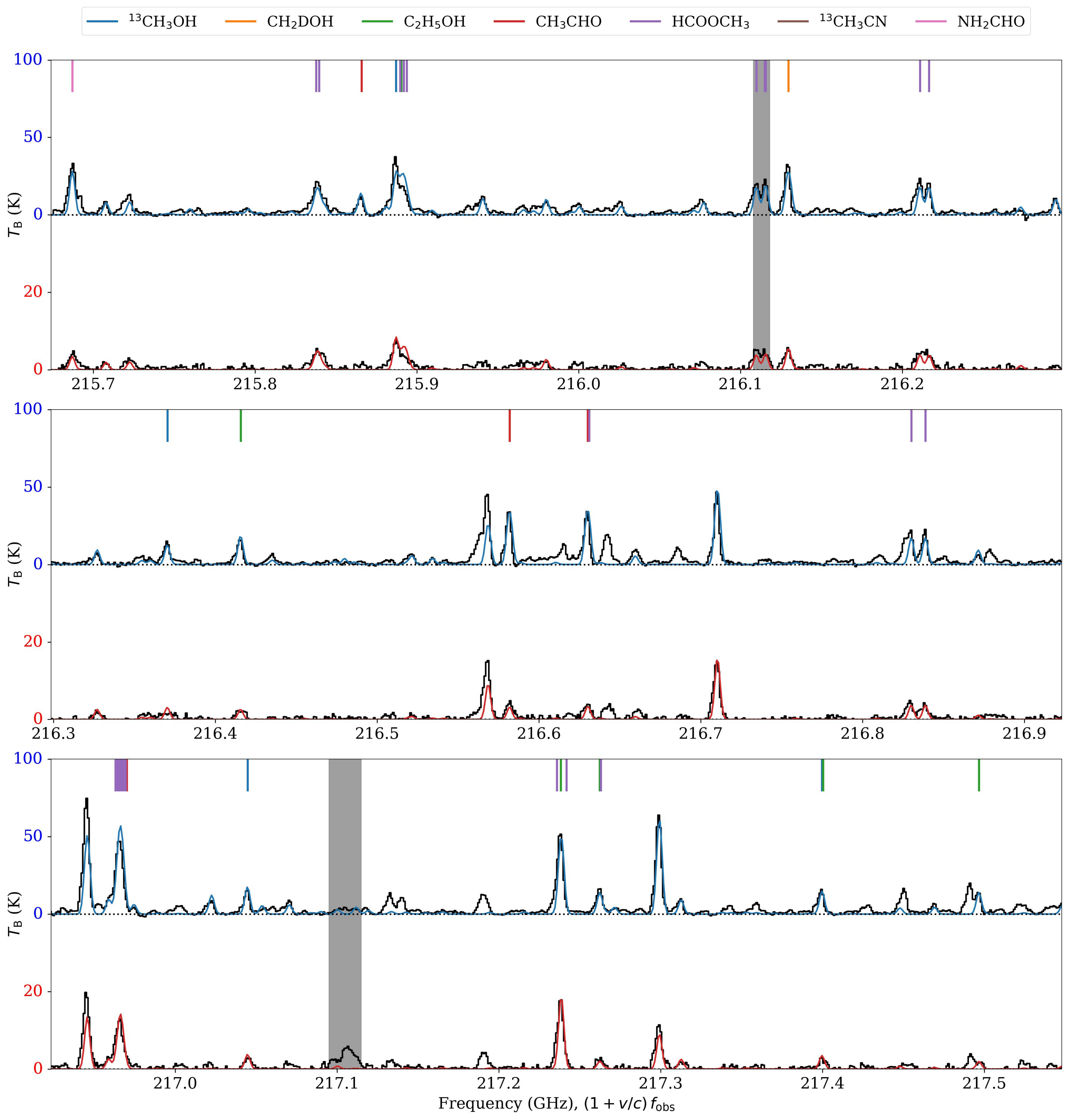}
\caption{\label{fig:appx_spec_spw3} 
Observed and synthesized spectra. 
Each panel is divided into two parts: the upper part corresponds to HOPS-288-A and the lower part to HOPS-288-B.
The black spectra show the observation results, while the blue and red curves represent the synthesized spectra for HOPS-288-A and HOPS-288-B, respectively.
Shaded regions indicate frequency ranges excluded from the fitting due to known transitions with non-Gaussian profiles, including CO, C$^{18}$O, SiO, and DCO$^+$.
Short vertical ticks at the top of each panel mark the selected COM transitions.
The x-axis shows frequency shifted to the expected rest frame using $v = 2.0$ km s$^{-1}$ and $v = 6.5$ km s$^{-1}$ for HOPS-288-A and HOPS-288-B, respectively.
}
\end{figure*}

\begin{figure*}[htb!]
\centering
\includegraphics[width=\linewidth]{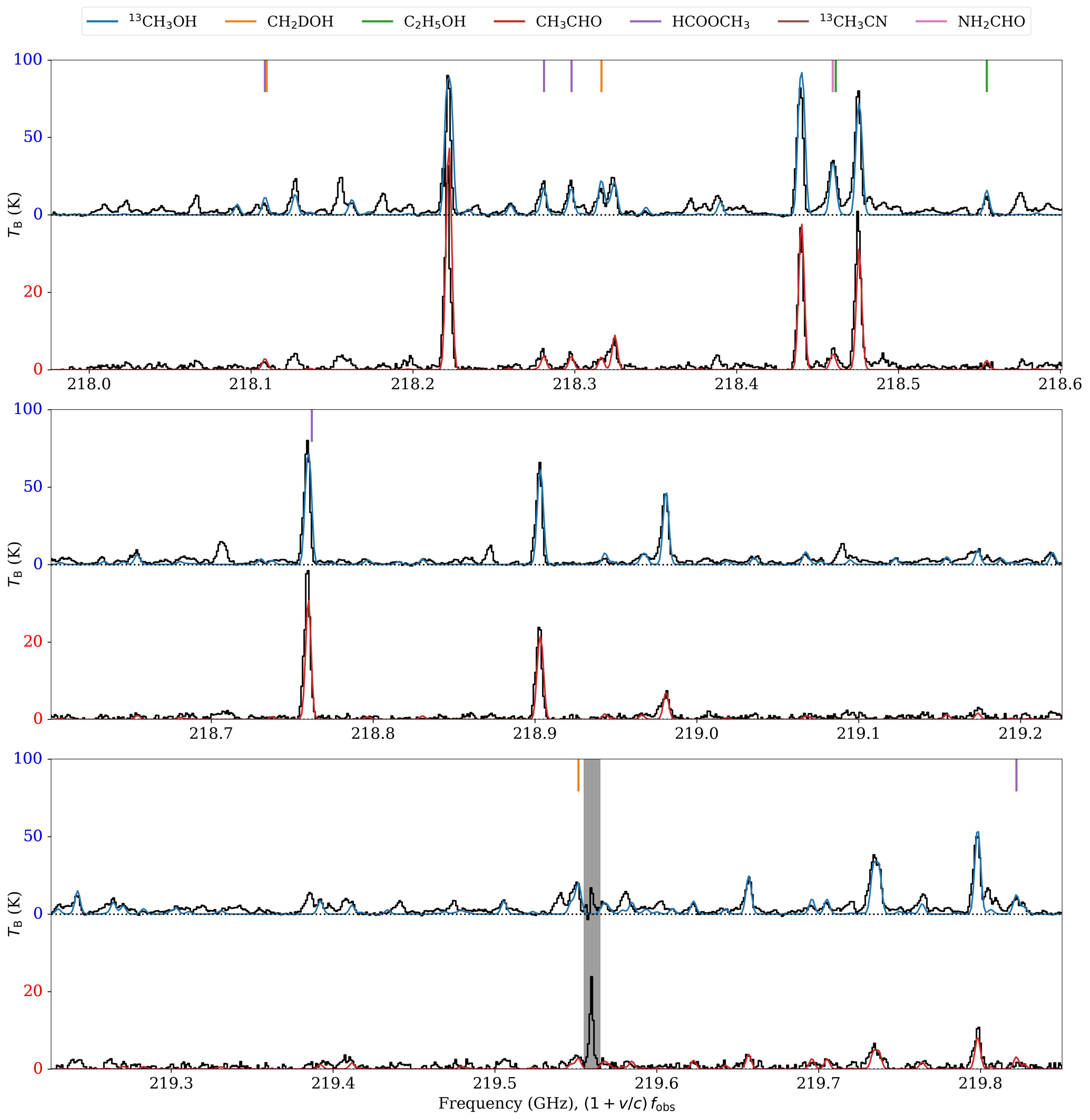}
\caption{\label{fig:appx_spec_spw2} 
The captions follow Figure~\ref{fig:appx_spec_spw3} with different frequency ranges. 
}
\end{figure*}

\begin{figure*}[htb!]
\centering
\includegraphics[width=\linewidth]{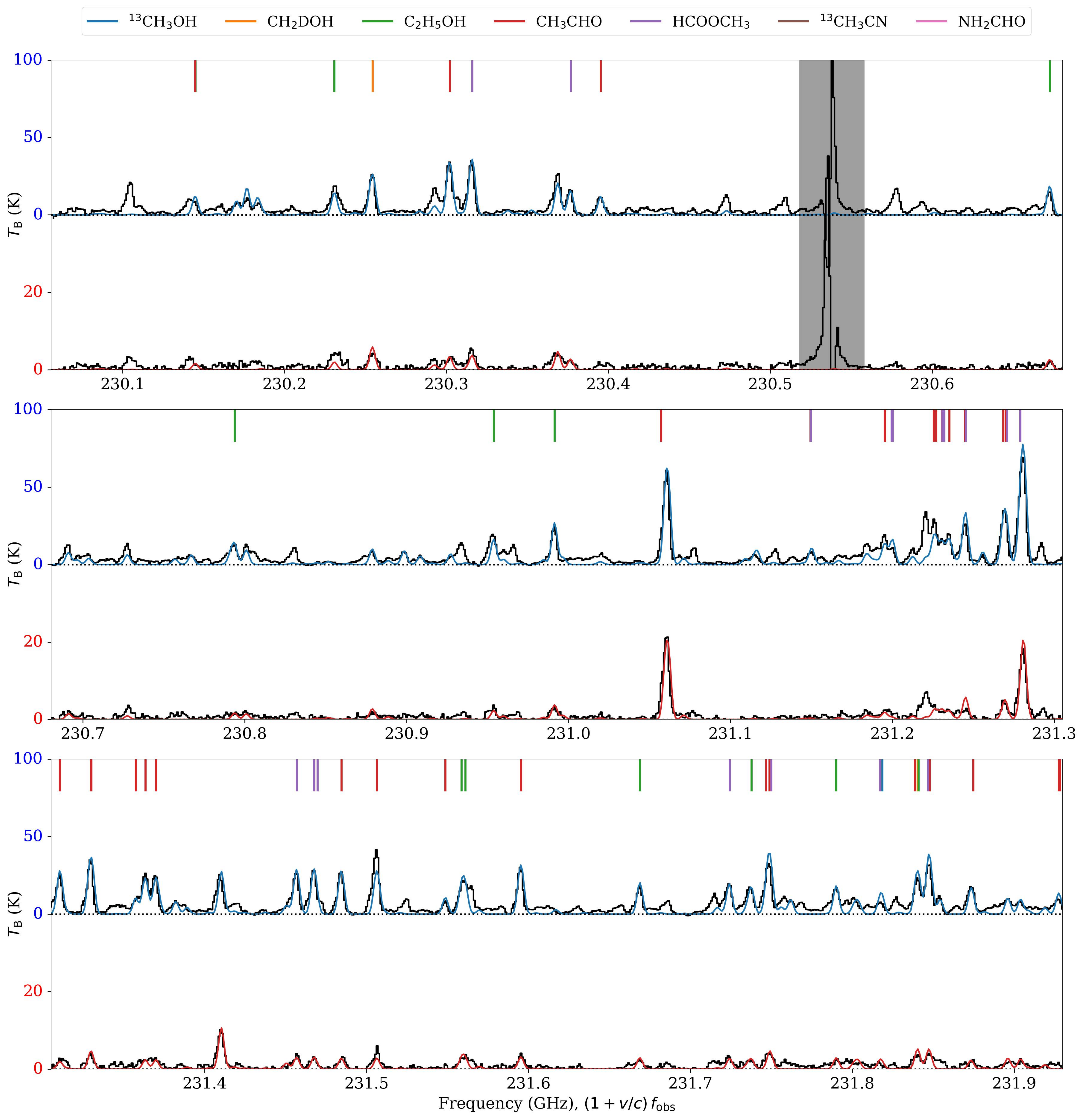}
\caption{\label{fig:appx_spec_spw0} 
The captions follow Figure~\ref{fig:appx_spec_spw3} with different frequency ranges. 
}
\end{figure*}

\begin{figure*}[htb!]
\centering
\includegraphics[width=\linewidth]{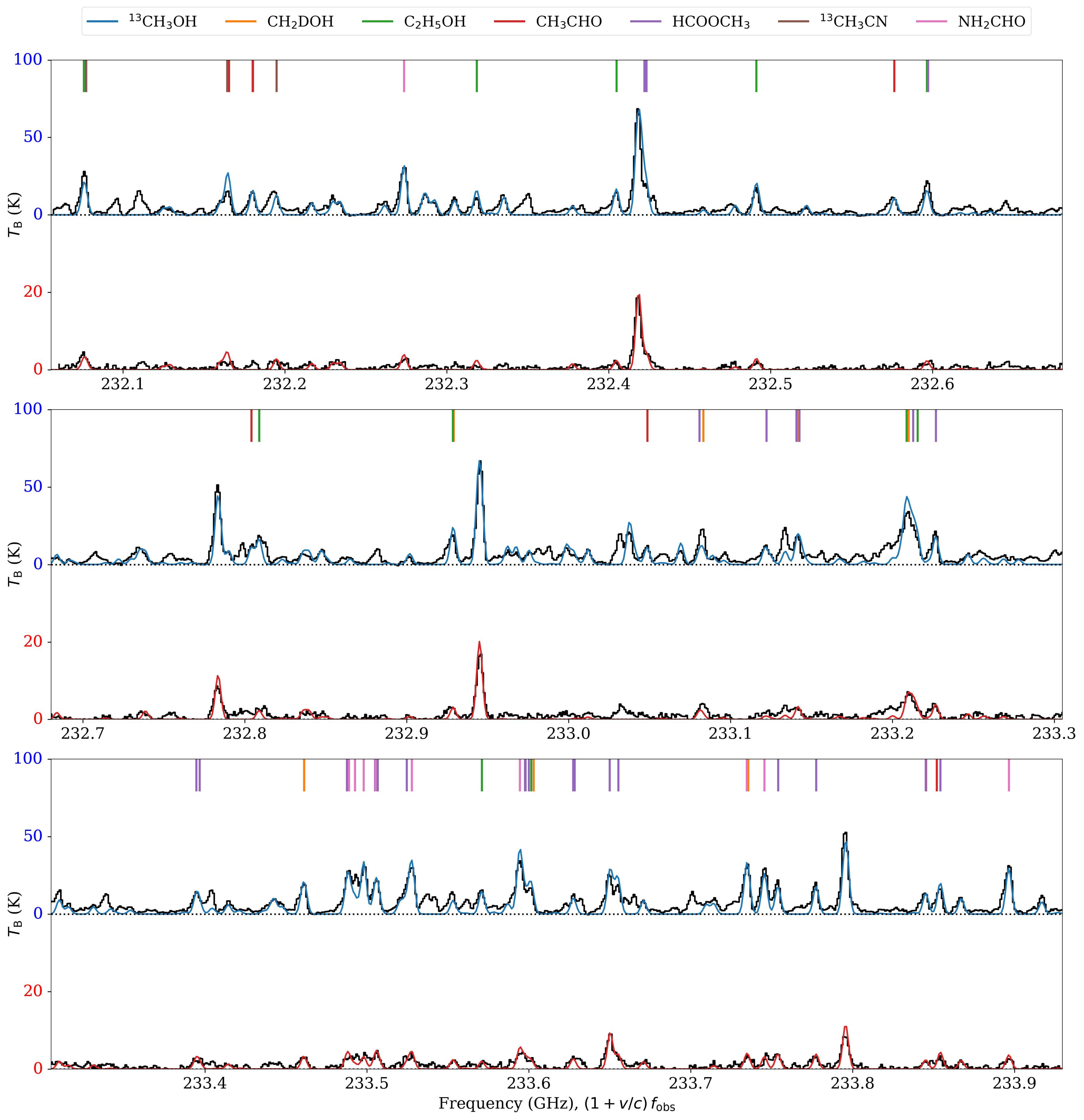}
\caption{\label{fig:appx_spec_spw1} 
The captions follow Figure~\ref{fig:appx_spec_spw3} with different frequency ranges. 
}
\end{figure*}

Tables~\ref{tab:appx:trans:Tobin} and \ref{tab:appx:trans:almasop} list the molecular transitions covering by the high-resolution (\#2018.1.01038.S) and wide-bandwidth (\#2018.1.00302.S), respectively. 
Figures~\ref{fig:appx_spec_spw3}, \ref{fig:appx_spec_spw2}, \ref{fig:appx_spec_spw0}, and \ref{fig:appx_spec_spw1} show the observed spectra and the modeled spectra from XCLASS fitting. 

\begin{deluxetable}{lrrrrl}
\tabletypesize{\scriptsize}
\setlength{\tabcolsep}{2pt}
\tablecaption{\label{tab:appx:trans:Tobin} Transitions used in this study covered by the high-resolution (\#2018.1.01038.S) data. }
\tablehead{
\colhead{Species} & \colhead{\frest} & \colhead{\Eu} & \colhead{\Aij} & \colhead{\gu} & \colhead{Quantum Numbers} \\
\colhead{} & \colhead{(MHz)} & \colhead{(K)} & \colhead{(s$^{-1}$)} & \colhead{} & \colhead{}
}
\startdata
CO & 230538.00 & 17 & 6.9106E-07 & 5 & v=0; $2-1$ \\
C$^{18}$O & 219560.35 & 16 & 6.0116E-07 & 5 & v=0; $2-1$ \\
SO & 219949.44 & 35 & 1.3352E-04 & 13 & S=1; v=0; J=6-5; N=5-4 \\
CH$_3$OH & 218440.05 & 45 & 4.6863E-05 & 36 & v=0; $4_{(2,3)}-3_{(1,2)}$; E \\
CH$_3$OH & 232945.00 & 190 & 2.1267E-05 & 84 & v=0; $10_{(3,7)}-11_{(2,9)}$; E \\
CH$_3$OH & 232783.50 & 447 & 2.1649E-05 & 148 & v=0; $18_{(3,16)}-17_{(4,13)}$; A1 \\
CH$_3$OH & 230368.70 & 683 & 2.0794E-05 & 180 & v=0; $22_{(4,18)}-21_{(5,16)}$; E \\
CH$_2$DOH & 233133.91 & 68 & 2.3468E-06 & 11 & v=0; $5_{(3,3)}-4_{(2,3)}$; e0-o1 \\
CH$_2$DOH & 233083.13 & 68 & 2.3254E-06 & 11 & v=0; $5_{(3,2)}-4_{(2,2)}$; e0-o1 \\
CH$_2$DOH & 233461.15 & 123 & 8.6800E-06 & 19 & v=0; $9_{(2,8)}-9_{(1,9)}$; e1 \\
CH$_3$CHO & 230301.92 & 81 & 4.1938E-04 & 50 & v=0; $12_{(2,11)}-11_{(2,10)}$; A \\
CH$_3$CHO & 230315.79 & 81 & 4.1933E-04 & 50 & v=0; $12_{(2,11)}-11_{(2,10)}$; E \\
HCOOCH$_3$ & 233777.52 & 114 & 1.8391E-04 & 74 & v=0; $18_{(4,14)}-17_{(4,13)}$; A \\
HCOOCH$_3$ & 233867.19 & 192 & 1.2860E-04 & 78 & v=0; $19_{(11,9)}-18_{(11,8)}$; E \\
C$_2$H$_5$OH & 230991.38 & 86 & 1.1962E-04 & 29 & $14_{(0,14)}-13_{(1,13)}$; A \\
C$_2$H$_5$OH & 230672.56 & 139 & 1.0602E-04 & 27 & $13_{(2,11)}-12_{(2,10)}$; G$^+$ \\
C$_2$H$_5$OH & 230230.74 & 143 & 8.2078E-05 & 27 & $13_{(2,11)}-12_{(2,10)}$; G$^-$ \\
C$_2$H$_5$OH & 230953.78 & 146 & 7.7648E-05 & 33 & $16_{(5,11)}-16_{(4,12)}$; A \\
C$_2$H$_5$OH & 230473.01 & 215 & 1.4244E-05 & 37 & $18_{(3,16)}-17_{(4,14)}$; G$^--$G$^+$ \\
C$_2$H$_5$CN & 233443.10 & 179 & 1.0368E-03 & 53 & $26_{(5,22)}-25_{(5,21)}$ \\
C$_2$H$_5$CN & 233069.31 & 205 & 9.9390E-04 & 53 & $26_{(7,20)}-25_{(7,19)}$ \\
C$_2$H$_5$CN & 233069.38 & 205 & 9.9390E-04 & 53 & $26_{(7,19)}-25_{(7,18)}$ \\
C$_2$H$_5$CN & 232998.74 & 222 & 9.6916E-04 & 53 & $26_{(8,19)}-25_{(8,18)}$ \\
C$_2$H$_5$CN & 232998.74 & 222 & 9.6916E-04 & 53 & $26_{(8,18)}-25_{(8,17)}$ \\
NH$_2$CHO & 233896.58 & 94 & 8.6203E-04 & 23 & $11_{(3,9)}-10_{(3,8)}$
\enddata
\end{deluxetable}

\startlongtable
\begin{deluxetable}{lrrrrl}
\tabletypesize{\scriptsize}
\tablecaption{\label{tab:appx:trans:almasop} Transitions used in this study covered by the wide-bandwidth (\#2018.1.00302.S) data. }
\setlength{\tabcolsep}{2pt}
\tablehead{
\colhead{Species} & \colhead{\frest} & \colhead{\Eu} & \colhead{\Aij} & \colhead{\gu} & \colhead{Quantum Numbers} \\
\colhead{} & \colhead{(MHz)} & \colhead{(K)} & \colhead{(s$^{-1}$)} & \colhead{} & \colhead{}
}
\startdata
H$_2$CO & 216568.65 & 174 & 7.2219E-06 & 57 & $9_{(1,8)}-9_{(1,9)}$ \\
H$_2$CO & 218760.07 & 68 & 1.5774E-04 & 7 & $3_{(2,1)}-2_{(2,0)}$ \\
H$_2$CO & 218475.63 & 68 & 1.5714E-04 & 7 & $3_{(2,2)}-2_{(2,1)}$ \\
H$_2$CO & 218222.19 & 21 & 2.8180E-04 & 7 & $3_{(0,3)}-2_{(0,2)}$ \\
D$_2$CO & 233650.44 & 50 & 2.6888E-04 & 18 & $4_{(2,3)}-3_{(2,2)}$ \\
D$_2$CO & 231410.23 & 28 & 3.4744E-04 & 18 & $4_{(0,4)}-3_{(0,3)}$ \\
CH$_3$OH & 218440.05 & 45 & 4.6863E-05 & 36 & v=0; $4_{(2,3)}-3_{(1,2)}$; E \\
CH$_3$OH & 216945.60 & 56 & 1.2135E-05 & 44 & v=0; $5_{(1,4)}-4_{(2,3)}$; E \\
CH$_3$OH & 232418.59 & 165 & 1.8675E-05 & 84 & v=0; $10_{(2,8)}-9_{(3,7)}$; A1 \\
CH$_3$OH & 231281.10 & 165 & 1.8314E-05 & 84 & v=0; $10_{(2,9)}-9_{(3,6)}$; A2 \\
CH$_3$OH & 232945.00 & 190 & 2.1267E-05 & 84 & v=0; $10_{(3,7)}-11_{(2,9)}$; E \\
CH$_3$OH & 233795.75 & 447 & 2.1978E-05 & 148 & v=0; $18_{(3,15)}-17_{(4,14)}$; A2 \\
CH$_3$OH & 232783.50 & 447 & 2.1649E-05 & 148 & v=0; $18_{(3,16)}-17_{(4,13)}$; A1 \\
CH$_3$OH & 230368.70 & 683 & 2.0794E-05 & 180 & v=0; $22_{(4,18)}-21_{(5,16)}$; E \\
CH$_3$OH & 233121.26 & 745 & 1.0539E-05 & 92 & v=1; $11_{(6,5)}-11_{(7,4)}$; E \\
CH$_3$OH & 217299.20 & 374 & 4.2846E-05 & 52 & v=1; $6_{(1,5)}-7_{(2,5)}$; A2 \\
$^{13}$CH$_3$OH & 233487.92 & 815 & 2.1510E-05 & 49 & v=0; $24_{(5,20)}-23_{(6,18)}$; E2 \\
$^{13}$CH$_3$OH & 231818.38 & 594 & 3.8732E-05 & 45 & v=0; $22_{(1,21)}-22_{(0,22)}$; E2 \\
$^{13}$CH$_3$OH & 217044.62 & 254 & 2.3746E-05 & 29 & v=0; $14_{(1,13)}-13_{(2,12)}$; A2 \\
$^{13}$CH$_3$OH & 217399.55 & 162 & 1.5272E-05 & 21 & v=0; $10_{(2,8)}-9_{(3,7)}$; A1 \\
$^{13}$CH$_3$OH & 216370.39 & 162 & 1.4987E-05 & 21 & v=0; $10_{(2,9)}-9_{(3,6)}$; A2 \\
$^{13}$CH$_3$OH & 215886.96 & 45 & 4.5276E-05 & 9 & v=0; $4_{(2,2)}-3_{(1,2)}$; E2 \\
CH$_2$DOH & 219551.49 & 48 & 5.0949E-06 & 11 & v=0; $5_{(1,5)}-4_{(1,4)}$; e1 \\
CH$_2$DOH & 233083.13 & 68 & 2.3254E-06 & 11 & v=0; $5_{(3,2)}-4_{(2,2)}$; e0-o1 \\
CH$_2$DOH & 218316.39 & 59 & 6.6880E-06 & 11 & v=0; $5_{(2,4)}-5_{(1,5)}$; e1 \\
CH$_2$DOH & 232929.02 & 49 & 3.2564E-06 & 7 & v=0; $3_{(3,0)}-4_{(2,3)}$; e0 \\
CH$_2$DOH & 233461.15 & 123 & 8.6800E-06 & 19 & v=0; $9_{(2,8)}-9_{(1,9)}$; e1 \\
CH$_2$DOH & 231840.30 & 124 & 1.2209E-05 & 21 & v=0; $10_{(1,9)}-9_{(2,8)}$; e0 \\
CH$_2$DOH & 216129.63 & 167 & 1.5496E-05 & 25 & v=0; $12_{(0,12)}-11_{(1,11)}$; e0-e1 \\
CH$_2$DOH & 233141.75 & 261 & 2.2225E-05 & 29 & v=0; $14_{(2,12)}-14_{(1,13)}$; o1 \\
CH$_2$DOH & 230376.57 & 293 & 2.2760E-05 & 31 & v=0; $15_{(2,13)}-15_{(1,14)}$; o1 \\
CH$_2$DOH & 230254.23 & 310 & 6.5037E-05 & 33 & v=0; $16_{(2,14)}-16_{(1,15)}$; e0 \\
CH$_2$DOH & 233209.88 & 347 & 6.4920E-05 & 35 & v=0; $17_{(2,15)}-17_{(1,16)}$; e0 \\
CH$_2$DOH & 233602.96 & 884 & 2.4365E-05 & 57 & v=0; $28_{(2,27)}-28_{(1,28)}$; e0-e1 \\
CH$_2$DOH & 233735.66 & 105 & 2.6980E-06 & 15 & v=0; $7_{(3,5)}-8_{(2,6)}$; e1-e0 \\
C$_2$H$_5$OH & 218554.51 & 226 & 7.1845E-05 & 43 & $21_{(5,16)}-21_{(4,17)}$;  A \\
C$_2$H$_5$OH & 231790.06 & 245 & 8.4615E-05 & 45 & $22_{(5,18)}-22_{(4,19)}$;  A \\
C$_2$H$_5$OH & 232318.51 & 264 & 8.5858E-05 & 47 & $23_{(5,19)}-23_{(4,20)}$;  A \\
C$_2$H$_5$OH & 230144.81 & 284 & 5.4589E-05 & 51 & $25_{(3,23)}-25_{(2,24)}$;  A \\
C$_2$H$_5$OH & 233208.54 & 284 & 8.7396E-05 & 49 & $24_{(5,20)}-24_{(4,21)}$;  A \\
C$_2$H$_5$OH & 217238.29 & 792 & 1.5185E-05 & 81 & $40_{(9,31)}-39_{(10,30)}$; A \\
C$_2$H$_5$OH & 217400.43 & 464 & 1.7655E-05 & 63 & $31_{(6,26)}-30_{(7,23)}$;  A \\
C$_2$H$_5$OH & 231789.85 & 506 & 2.2387E-05 & 63 & $31_{(5,27)}-31_{(4,27)}$;  G$^--$G$^+$ \\
C$_2$H$_5$OH & 231840.75 & 685 & 1.0785E-05 & 71 & $35_{(9,26)}-34_{(10,24)}$; G$^--$G$^+$ \\
C$_2$H$_5$OH & 232422.96 & 923 & 1.5956E-05 & 87 & $43_{(8,35)}-42_{(9,33)}$;  G$^+-$G$^-$ \\
C$_2$H$_5$OH & 215890.45 & 294 & 6.3232E-05 & 51 & $25_{(4,22)}-25_{(3,23)}$;  A \\
C$_2$H$_5$OH & 231558.57 & 226 & 8.3570E-05 & 43 & $21_{(5,17)}-21_{(4,18)}$;  A \\
C$_2$H$_5$OH & 231737.62 & 191 & 8.1839E-05 & 39 & $19_{(5,15)}-19_{(4,16)}$;  A \\
C$_2$H$_5$OH & 230793.86 & 105 & 6.2028E-05 & 13 & $6_{(5,2)}-5_{(4,2)}$;      G$^+-$G$^-$ \\
C$_2$H$_5$OH & 230793.76 & 105 & 6.2028E-05 & 13 & $6_{(5,1)}-5_{(4,1)}$;      G$^+-$G$^-$ \\
C$_2$H$_5$OH & 233601.55 & 120 & 7.7124E-05 & 29 & $14_{(5,10)}-14_{(4,11)}$;  A \\
C$_2$H$_5$OH & 232928.53 & 120 & 7.6535E-05 & 29 & $14_{(5,9)}-14_{(4,10)}$;   A \\
C$_2$H$_5$OH & 216415.67 & 131 & 9.0183E-05 & 27 & $13_{(0,13)}-12_{(0,12)}$;  G$^+$ \\
C$_2$H$_5$OH & 233215.50 & 132 & 7.8327E-05 & 31 & $15_{(5,11)}-15_{(4,12)}$;  A \\
C$_2$H$_5$OH & 230991.38 & 86 & 1.1962E-04 & 29 & $14_{(0,14)}-13_{(1,13)}$;   A \\
C$_2$H$_5$OH & 218461.23 & 24 & 6.3782E-05 & 11 & $5_{(3,2)}-4_{(2,3)}$;       A \\
C$_2$H$_5$OH & 231560.92 & 208 & 8.2653E-05 & 41 & $20_{(5,16)}-20_{(4,17)}$;  A \\
C$_2$H$_5$OH & 217496.67 & 135 & 6.8662E-05 & 27 & $13_{(1,13)}-12_{(0,12)}$;  G$^-$ \\
C$_2$H$_5$OH & 232075.85 & 132 & 7.7291E-05 & 31 & $15_{(5,10)}-15_{(4,11)}$;  A \\
C$_2$H$_5$OH & 230672.56 & 139 & 1.0602E-04 & 27 & $13_{(2,11)}-12_{(2,10)}$;  G$^+$ \\
C$_2$H$_5$OH & 232491.32 & 142 & 1.1210E-04 & 29 & $14_{(0,14)}-13_{(0,13)}$;  G$^+$ \\
C$_2$H$_5$OH & 231668.74 & 142 & 1.1080E-04 & 29 & $14_{(1,14)}-13_{(1,13)}$;  G$^+$ \\
C$_2$H$_5$OH & 230230.74 & 143 & 8.2078E-05 & 27 & $13_{(2,11)}-12_{(2,10)}$;  G$^-$ \\
C$_2$H$_5$OH & 232808.86 & 146 & 7.9330E-05 & 33 & $16_{(5,12)}-16_{(4,13)}$;  A \\
C$_2$H$_5$OH & 230953.78 & 146 & 7.7648E-05 & 33 & $16_{(5,11)}-16_{(4,12)}$;  A \\
C$_2$H$_5$OH & 232596.57 & 147 & 8.0717E-05 & 29 & $14_{(1,14)}-13_{(1,13)}$;  G$^-$ \\
C$_2$H$_5$OH & 232404.84 & 160 & 8.0221E-05 & 35 & $17_{(5,13)}-17_{(4,14)}$;  A \\
C$_2$H$_5$OH & 217262.30 & 136 & 6.2719E-05 & 27 & $13_{(0,13)}-12_{(1,12)}$;  G$^-$ \\
C$_2$H$_5$OH & 230793.86 & 105 & 6.2028E-05 & 13 & $6_{(5,2)}-5_{(4,2)}$;      G$^+-$G$^-$ \\
C$_2$H$_5$OH & 230793.76 & 105 & 6.2028E-05 & 13 & $6_{(5,1)}-5_{(4,1)}$;      G$^+-$G$^-$ \\
CH$_3$CHO & 231484.37 & 108 & 3.8971E-04 & 50 & v=0; $12_{(4,8)}-11_{(4,7)}$; E \\
CH$_3$CHO & 231506.29 & 108 & 3.8969E-04 & 50 & v=0; $12_{(4,9)}-11_{(4,8)}$; E \\
CH$_3$CHO & 231748.72 & 93 & 4.0896E-04 & 50 & v=0; $12_{(3,10)}-11_{(3,9)}$; E \\
CH$_3$CHO & 231595.27 & 93 & 4.1153E-04 & 50 & v=0; $12_{(3,10)}-11_{(3,9)}$; A \\
CH$_3$CHO & 231225.54 & 183 & 2.8833E-04 & 50 & v=0; $12_{(7,5)}-11_{(7,4)}$; E \\
CH$_3$CHO & 230315.79 & 81 & 4.1933E-04 & 50 & v=0; $12_{(2,11)}-11_{(2,10)}$; E \\
CH$_3$CHO & 230301.92 & 81 & 4.1938E-04 & 50 & v=0; $12_{(2,11)}-11_{(2,10)}$; A \\
CH$_3$CHO & 216581.93 & 65 & 3.5458E-04 & 46 & v=0; $11_{(1,10)}-10_{(1,9)}$; E \\
CH$_3$CHO & 231847.58 & 93 & 4.0964E-04 & 50 & v=0; $12_{(3,9)}-11_{(3,8)}$; E \\
CH$_3$CHO & 231245.03 & 183 & 2.8843E-04 & 50 & v=0; $12_{(7,5)}-11_{(7,4)}$; A \\
CH$_3$CHO & 231245.03 & 183 & 2.8843E-04 & 50 & v=0; $12_{(7,6)}-11_{(7,5)}$; A \\
CH$_3$CHO & 231268.39 & 183 & 2.8842E-04 & 50 & v=0; $12_{(7,6)}-11_{(7,5)}$; E \\
CH$_3$CHO & 231269.90 & 153 & 3.2800E-04 & 50 & v=0; $12_{(6,7)}-11_{(6,6)}$; A \\
CH$_3$CHO & 231269.90 & 153 & 3.2800E-04 & 50 & v=0; $12_{(6,6)}-11_{(6,5)}$; A \\
CH$_3$CHO & 231278.98 & 153 & 3.2793E-04 & 50 & v=0; $12_{(6,6)}-11_{(6,5)}$; E \\
CH$_3$CHO & 231310.50 & 153 & 3.2803E-04 & 50 & v=0; $12_{(6,7)}-11_{(6,6)}$; E \\
CH$_3$CHO & 231456.74 & 108 & 3.8965E-04 & 50 & v=0; $12_{(4,9)}-11_{(4,8)}$; A \\
CH$_3$CHO & 231467.50 & 108 & 3.8967E-04 & 50 & v=0; $12_{(4,8)}-11_{(4,7)}$; A \\
CH$_3$CHO & 231369.83 & 129 & 3.6174E-04 & 50 & v=0; $12_{(5,8)}-11_{(5,7)}$; E \\
CH$_3$CHO & 231363.28 & 129 & 3.6168E-04 & 50 & v=0; $12_{(5,7)}-11_{(5,6)}$; E \\
CH$_3$CHO & 231329.64 & 129 & 3.6172E-04 & 50 & v=0; $12_{(5,8)}-11_{(5,7)}$; A \\
CH$_3$CHO & 231329.79 & 129 & 3.6172E-04 & 50 & v=0; $12_{(5,7)}-11_{(5,6)}$; A \\
CH$_3$CHO & 231245.03 & 183 & 2.8843E-04 & 50 & v=0; $12_{(7,5)}-11_{(7,4)}$; A \\
CH$_3$CHO & 231245.03 & 183 & 2.8843E-04 & 50 & v=0; $12_{(7,6)}-11_{(7,5)}$; A \\
CH$_3$CHO & 231269.90 & 153 & 3.2800E-04 & 50 & v=0; $12_{(6,7)}-11_{(6,6)}$; A \\
CH$_3$CHO & 231269.90 & 153 & 3.2800E-04 & 50 & v=0; $12_{(6,6)}-11_{(6,5)}$; A \\
CH$_3$CHO & 231235.14 & 216 & 2.4279E-04 & 50 & v=0; $12_{(8,4)}-11_{(8,3)}$; A \\
CH$_3$CHO & 231235.14 & 216 & 2.4279E-04 & 50 & v=0; $12_{(8,5)}-11_{(8,4)}$; A \\
CH$_3$CHO & 216969.89 & 556 & 1.0525E-05 & 134 & v=0; $33_{(4,30)}-32_{(5,27)}$; E \\
CH$_3$CHO & 231226.97 & 255 & 1.9119E-04 & 50 & v=0; $12_{(9,3)}-11_{(9,2)}$; A \\
CH$_3$CHO & 231226.97 & 255 & 1.9119E-04 & 50 & v=0; $12_{(9,4)}-11_{(9,3)}$; A \\
CH$_3$CHO & 231226.97 & 255 & 1.9119E-04 & 50 & v=0; $12_{(9,3)}-11_{(9,2)}$; A \\
CH$_3$CHO & 231226.97 & 255 & 1.9119E-04 & 50 & v=0; $12_{(9,4)}-11_{(9,3)}$; A \\
CH$_3$CHO & 231357.35 & 299 & 4.1129E-04 & 50 & v=1; $12_{(3,9)}-11_{(3,8)}$; E \\
CH$_3$CHO & 233048.52 & 285 & 4.3142E-04 & 50 & v=1; $12_{(2,11)}-11_{(2,10)}$; E \\
CH$_3$CHO & 232576.45 & 297 & 4.1627E-04 & 50 & v=1; $12_{(3,9)}-11_{(3,8)}$; A \\
CH$_3$CHO & 230395.16 & 286 & 4.2196E-04 & 50 & v=1; $12_{(2,11)}-11_{(2,10)}$; A \\
CH$_3$CHO & 232804.10 & 288 & 4.3087E-04 & 50 & v=1; $12_{(2,10)}-11_{(2,9)}$; E \\
CH$_3$CHO & 230144.55 & 290 & 4.8749E-05 & 54 & v=1; $13_{(0,13)}-12_{(1,12)}$; A \\
CH$_3$CHO & 232422.03 & 297 & 4.1335E-04 & 50 & v=1; $12_{(3,10)}-11_{(3,9)}$; E \\
CH$_3$CHO & 232165.52 & 297 & 4.1414E-04 & 50 & v=1; $12_{(3,10)}-11_{(3,9)}$; A \\
CH$_3$CHO & 233852.03 & 784 & 5.4483E-05 & 138 & v=1; $34_{(3,31)}-34_{(2,32)}$; A \\
CH$_3$CHO & 231927.65 & 763 & 1.0281E-05 & 114 & v=1; $28_{(9,19)}-29_{(8,22)}$; A \\
CH$_3$CHO & 231928.14 & 763 & 1.0281E-05 & 114 & v=1; $28_{(9,20)}-29_{(8,21)}$; A \\
CH$_3$CHO & 231874.70 & 457 & 1.9107E-04 & 50 & v=1; $12_{(9,4)}-11_{(9,3)}$; E \\
CH$_3$CHO & 231927.65 & 763 & 1.0281E-05 & 114 & v=1; $28_{(9,19)}-29_{(8,22)}$; A \\
CH$_3$CHO & 231928.14 & 763 & 1.0281E-05 & 114 & v=1; $28_{(9,20)}-29_{(8,21)}$; A \\
CH$_3$CHO & 231927.47 & 357 & 3.2852E-04 & 50 & v=1; $12_{(6,7)}-11_{(6,6)}$; A \\
CH$_3$CHO & 231927.47 & 357 & 3.2852E-04 & 50 & v=1; $12_{(6,6)}-11_{(6,5)}$; A \\
CH$_3$CHO & 232180.10 & 332 & 3.6284E-04 & 50 & v=1; $12_{(5,8)}-11_{(5,7)}$; A \\
CH$_3$CHO & 232180.27 & 332 & 3.6284E-04 & 50 & v=1; $12_{(5,7)}-11_{(5,6)}$; A \\
CH$_3$CHO & 231548.63 & 387 & 2.8816E-04 & 50 & v=1; $12_{(7,6)}-11_{(7,5)}$; A \\
CH$_3$CHO & 231548.63 & 387 & 2.8816E-04 & 50 & v=1; $12_{(7,5)}-11_{(7,4)}$; A \\
CH$_3$CHO & 231746.68 & 357 & 3.2869E-04 & 50 & v=1; $12_{(6,7)}-11_{(6,6)}$; E \\
HCOOCH$_3$ & 216838.89 & 106 & 1.4799E-04 & 74 & v=0; $18_{(2,16)}-17_{(2,15)}$; A \\
HCOOCH$_3$ & 216967.42 & 111 & 1.5319E-04 & 82 & v=0; $20_{(0,20)}-19_{(0,19)}$; A \\
HCOOCH$_3$ & 216830.20 & 106 & 1.4796E-04 & 74 & v=0; $18_{(2,16)}-17_{(2,15)}$; E \\
HCOOCH$_3$ & 216216.54 & 109 & 1.4898E-04 & 78 & v=0; $19_{(1,18)}-18_{(1,17)}$; A \\
HCOOCH$_3$ & 216115.57 & 109 & 1.4875E-04 & 78 & v=0; $19_{(2,18)}-18_{(2,17)}$; A \\
HCOOCH$_3$ & 216210.91 & 109 & 1.4895E-04 & 78 & v=0; $19_{(1,18)}-18_{(1,17)}$; E \\
HCOOCH$_3$ & 216109.78 & 109 & 1.4872E-04 & 78 & v=0; $19_{(2,18)}-18_{(2,17)}$; E \\
HCOOCH$_3$ & 216969.19 & 111 & 2.4436E-05 & 82 & v=0; $20_{(1,20)}-19_{(0,19)}$; A \\
HCOOCH$_3$ & 216962.99 & 111 & 2.4448E-05 & 82 & v=0; $20_{(0,20)}-19_{(1,19)}$; E \\
HCOOCH$_3$ & 216965.90 & 111 & 1.5315E-04 & 82 & v=0; $20_{(1,20)}-19_{(1,19)}$; A \\
HCOOCH$_3$ & 216966.25 & 111 & 1.5313E-04 & 82 & v=0; $20_{(0,20)}-19_{(0,19)}$; E \\
HCOOCH$_3$ & 216967.99 & 111 & 2.4449E-05 & 82 & v=0; $20_{(1,20)}-19_{(0,19)}$; E \\
HCOOCH$_3$ & 216964.77 & 111 & 1.5313E-04 & 82 & v=0; $20_{(1,20)}-19_{(1,19)}$; E \\
HCOOCH$_3$ & 233777.52 & 114 & 1.8391E-04 & 74 & v=0; $18_{(4,14)}-17_{(4,13)}$; A \\
HCOOCH$_3$ & 233753.96 & 114 & 1.8385E-04 & 74 & v=0; $18_{(4,14)}-17_{(4,13)}$; E \\
HCOOCH$_3$ & 233226.79 & 123 & 1.8189E-04 & 78 & v=0; $19_{(4,16)}-18_{(4,15)}$; A \\
HCOOCH$_3$ & 233628.46 & 144 & 1.4443E-05 & 70 & v=0; $17_{(9,9)}-17_{(8,10)}$; A \\
HCOOCH$_3$ & 233627.48 & 144 & 1.4443E-05 & 70 & v=0; $17_{(9,8)}-17_{(8,9)}$; A \\
HCOOCH$_3$ & 233212.77 & 123 & 1.8184E-04 & 78 & v=0; $19_{(4,16)}-18_{(4,15)}$; E \\
HCOOCH$_3$ & 230315.80 & 203 & 1.5394E-05 & 90 & v=0; $22_{(9,14)}-22_{(8,15)}$; E \\
HCOOCH$_3$ & 233597.70 & 144 & 1.4442E-05 & 70 & v=0; $17_{(9,8)}-17_{(8,9)}$; E \\
HCOOCH$_3$ & 217235.92 & 368 & 1.5932E-05 & 130 & v=0; $32_{(9,24)}-32_{(8,25)}$; E \\
HCOOCH$_3$ & 216114.96 & 312 & 1.5134E-05 & 118 & v=0; $29_{(9,20)}-29_{(8,21)}$; A \\
HCOOCH$_3$ & 233212.77 & 304 & 3.8337E-05 & 78 & v=0; $19_{(17,2)}-18_{(17,1)}$; E \\
HCOOCH$_3$ & 231232.06 & 264 & 1.0043E-05 & 118 & v=0; $29_{(4,26)}-29_{(3,27)}$; E \\
HCOOCH$_3$ & 233394.65 & 242 & 8.7967E-05 & 78 & v=0; $19_{(14,5)}-18_{(14,4)}$; A \\
HCOOCH$_3$ & 233394.65 & 242 & 8.7967E-05 & 78 & v=0; $19_{(14,6)}-18_{(14,5)}$; A \\
HCOOCH$_3$ & 233396.68 & 242 & 8.7947E-05 & 78 & v=0; $19_{(14,5)}-18_{(14,4)}$; E \\
HCOOCH$_3$ & 233504.98 & 224 & 1.0245E-04 & 78 & v=0; $19_{(13,6)}-18_{(13,5)}$; E \\
HCOOCH$_3$ & 233506.69 & 224 & 1.0247E-04 & 78 & v=0; $19_{(13,7)}-18_{(13,6)}$; A \\
HCOOCH$_3$ & 233506.69 & 224 & 1.0247E-04 & 78 & v=0; $19_{(13,6)}-18_{(13,5)}$; A \\
HCOOCH$_3$ & 233524.63 & 224 & 1.0250E-04 & 78 & v=0; $19_{(13,7)}-18_{(13,6)}$; E \\
HCOOCH$_3$ & 233628.46 & 144 & 1.4443E-05 & 70 & v=0; $17_{(9,9)}-17_{(8,10)}$; A \\
HCOOCH$_3$ & 233627.48 & 144 & 1.4443E-05 & 70 & v=0; $17_{(9,8)}-17_{(8,9)}$; A \\
HCOOCH$_3$ & 233655.34 & 208 & 1.1599E-04 & 78 & v=0; $19_{(12,8)}-18_{(12,7)}$; A \\
HCOOCH$_3$ & 233655.34 & 208 & 1.1599E-04 & 78 & v=0; $19_{(12,7)}-18_{(12,6)}$; A \\
HCOOCH$_3$ & 218280.90 & 100 & 1.5077E-04 & 70 & v=0; $17_{(3,14)}-16_{(3,13)}$; E \\
HCOOCH$_3$ & 233845.23 & 192 & 1.2856E-04 & 78 & v=0; $19_{(11,8)}-18_{(11,7)}$; E \\
HCOOCH$_3$ & 233854.29 & 192 & 1.2860E-04 & 78 & v=0; $19_{(11,8)}-18_{(11,7)}$; A \\
HCOOCH$_3$ & 233854.29 & 192 & 1.2860E-04 & 78 & v=0; $19_{(11,9)}-18_{(11,8)}$; A \\
HCOOCH$_3$ & 231199.35 & 190 & 1.5875E-05 & 86 & v=0; $21_{(9,12)}-21_{(8,13)}$; A \\
HCOOCH$_3$ & 231200.14 & 190 & 1.5737E-05 & 86 & v=0; $21_{(9,12)}-21_{(8,13)}$; E \\
HCOOCH$_3$ & 232597.28 & 166 & 1.5301E-05 & 78 & v=0; $19_{(9,10)}-19_{(8,11)}$; E \\
HCOOCH$_3$ & 233140.74 & 155 & 1.4913E-05 & 74 & v=0; $18_{(9,9)}-18_{(8,10)}$; E \\
HCOOCH$_3$ & 233122.16 & 155 & 1.4915E-05 & 74 & v=0; $18_{(9,10)}-18_{(8,11)}$; E \\
HCOOCH$_3$ & 233655.34 & 208 & 1.1599E-04 & 78 & v=0; $19_{(12,8)}-18_{(12,7)}$; A \\
HCOOCH$_3$ & 233655.34 & 208 & 1.1599E-04 & 78 & v=0; $19_{(12,7)}-18_{(12,6)}$; A \\
HCOOCH$_3$ & 216631.08 & 440 & 1.5342E-05 & 146 & v=0; $36_{(8,29)}-36_{(7,30)}$; E \\
HCOOCH$_3$ & 231456.84 & 532 & 1.7389E-05 & 162 & v=0; $40_{(7,33)}-40_{(6,34)}$; A \\
HCOOCH$_3$ & 217262.88 & 485 & 1.6574E-05 & 150 & v=0; $37_{(10,27)}-37_{(9,28)}$; A \\
HCOOCH$_3$ & 231469.66 & 441 & 1.9260E-05 & 142 & v=0; $35_{(10,25)}-35_{(9,26)}$; E \\
HCOOCH$_3$ & 219822.13 & 355 & 1.1074E-04 & 74 & v=1; $18_{(10,9)}-17_{(10,8)}$; A \\
HCOOCH$_3$ & 219822.13 & 355 & 1.1074E-04 & 74 & v=1; $18_{(10,8)}-17_{(10,7)}$; A \\
HCOOCH$_3$ & 232164.44 & 366 & 1.3664E-04 & 78 & v=1; $19_{(10,9)}-18_{(10,8)}$; A \\
HCOOCH$_3$ & 232164.44 & 366 & 1.3664E-04 & 78 & v=1; $19_{(10,10)}-18_{(10,9)}$; A \\
HCOOCH$_3$ & 231749.76 & 366 & 1.3589E-04 & 78 & v=1; $19_{(10,9)}-18_{(10,8)}$; E \\
HCOOCH$_3$ & 231230.68 & 396 & 1.1221E-04 & 78 & v=1; $19_{(12,7)}-18_{(12,6)}$; E \\
HCOOCH$_3$ & 232422.05 & 404 & 1.6823E-05 & 94 & v=1; $23_{(9,15)}-23_{(8,16)}$; E \\
HCOOCH$_3$ & 232077.29 & 404 & 1.6897E-05 & 94 & v=1; $23_{(9,15)}-23_{(8,16)}$; A \\
HCOOCH$_3$ & 231846.82 & 412 & 1.0029E-04 & 78 & v=1; $19_{(13,6)}-18_{(13,5)}$; A \\
HCOOCH$_3$ & 231846.82 & 412 & 1.0029E-04 & 78 & v=1; $19_{(13,7)}-18_{(13,6)}$; A \\
HCOOCH$_3$ & 231270.76 & 418 & 1.6910E-05 & 98 & v=1; $24_{(9,16)}-24_{(8,17)}$; E \\
HCOOCH$_3$ & 231816.99 & 430 & 8.6209E-05 & 78 & v=1; $19_{(14,6)}-18_{(14,5)}$; A \\
HCOOCH$_3$ & 231816.99 & 430 & 8.6209E-05 & 78 & v=1; $19_{(14,5)}-18_{(14,4)}$; A \\
HCOOCH$_3$ & 217241.80 & 477 & 1.0797E-05 & 122 & v=1; $30_{(5,26)}-30_{(4,27)}$; E \\
HCOOCH$_3$ & 231278.96 & 493 & 3.7401E-05 & 78 & v=1; $19_{(17,3)}-18_{(17,2)}$; E \\
HCOOCH$_3$ & 231149.58 & 516 & 1.9188E-05 & 78 & v=1; $19_{(18,2)}-18_{(18,1)}$; E \\
HCOOCH$_3$ & 231467.34 & 562 & 1.4832E-05 & 138 & v=1; $34_{(6,29)}-34_{(5,30)}$; A \\
HCOOCH$_3$ & 233600.06 & 647 & 1.9970E-05 & 146 & v=1; $36_{(10,26)}-36_{(9,27)}$; A \\
HCOOCH$_3$ & 233142.48 & 949 & 1.9936E-05 & 194 & v=1; $48_{(9,39)}-48_{(8,40)}$; A \\
HCOOCH$_3$ & 218762.13 & 1050 & 1.7906E-05 & 206 & v=1; $51_{(10,41)}-51_{(9,42)}$; A \\
HCOOCH$_3$ & 232423.45 & 353 & 1.4701E-04 & 78 & v=1; $19_{(9,11)}-18_{(9,10)}$; A \\
HCOOCH$_3$ & 232423.45 & 353 & 1.4701E-04 & 78 & v=1; $19_{(9,10)}-18_{(9,9)}$; A \\
HCOOCH$_3$ & 233080.84 & 353 & 1.4869E-04 & 78 & v=1; $19_{(9,11)}-18_{(9,10)}$; E \\
HCOOCH$_3$ & 233627.48 & 341 & 1.5875E-04 & 78 & v=1; $19_{(8,12)}-18_{(8,11)}$; E \\
HCOOCH$_3$ & 233598.10 & 332 & 1.6650E-04 & 78 & v=1; $19_{(7,12)}-18_{(7,11)}$; E \\
HCOOCH$_3$ & 233487.68 & 332 & 1.6589E-04 & 78 & v=1; $19_{(7,13)}-18_{(7,12)}$; A \\
HCOOCH$_3$ & 231245.42 & 310 & 1.7698E-04 & 78 & v=1; $19_{(4,16)}-18_{(4,15)}$; A \\
HCOOCH$_3$ & 231724.16 & 301 & 1.7932E-04 & 74 & v=1; $18_{(4,14)}-17_{(4,13)}$; E \\
HCOOCH$_3$ & 215837.59 & 299 & 1.5046E-04 & 82 & v=1; $20_{(1,20)}-19_{(1,19)}$; A \\
HCOOCH$_3$ & 215839.54 & 299 & 1.5046E-04 & 82 & v=1; $20_{(0,20)}-19_{(0,19)}$; A \\
HCOOCH$_3$ & 215891.90 & 298 & 1.5126E-04 & 82 & v=1; $20_{(1,20)}-19_{(1,19)}$; E \\
HCOOCH$_3$ & 215889.66 & 298 & 2.3710E-05 & 82 & v=1; $20_{(0,20)}-19_{(1,19)}$; E \\
HCOOCH$_3$ & 215893.71 & 298 & 1.5126E-04 & 82 & v=1; $20_{(0,20)}-19_{(0,19)}$; E \\
CH$_2$(OH)CHO & 231724.33 & 42 & 2.5438E-04 & 17 & $8_{(6,2)}-7_{(5,3)}$ \\
CH$_2$(OH)CHO & 231723.27 & 42 & 2.5438E-04 & 17 & $8_{(6,3)}-7_{(5,2)}$ \\
CH$_2$(OH)CHO & 233797.25 & 95 & 1.3045E-04 & 35 & $17_{(4,14)}-16_{(3,13)}$ \\
CH$_2$(OH)CHO & 232335.45 & 135 & 3.0669E-04 & 45 & $22_{(2,21)}-21_{(1,20)}$ \\
CH$_2$(OH)CHO & 232286.03 & 135 & 3.0656E-04 & 45 & $22_{(1,21)}-21_{(2,20)}$ \\
CH$_2$(OH)CHO & 233037.73 & 137 & 3.6542E-04 & 47 & $23_{(1,23)}-22_{(0,22)}$ \\
CH$_2$(OH)CHO & 233037.36 & 137 & 3.6542E-04 & 47 & $23_{(0,23)}-22_{(1,22)}$ \\
CH$_2$(OH)CHO & 232288.41 & 271 & 2.0576E-04 & 55 & $27_{(10,17)}-27_{(9,18)}$ \\
CH$_2$(OH)CHO & 231200.38 & 287 & 2.0649E-04 & 57 & $28_{(10,19)}-28_{(9,20)}$ \\
CH$_2$(OH)CHO & 232782.05 & 804 & 4.7057E-05 & 99 & $49_{(14,35)}-48_{(15,34)}$ \\
CH$_2$(OH)CHO & 233037.73 & 137 & 3.6542E-04 & 47 & $23_{(1,23)}-22_{(0,22)}$ \\
CH$_2$(OH)CHO & 233037.36 & 137 & 3.6542E-04 & 47 & $23_{(0,23)}-22_{(1,22)}$ \\
CH$_3$COCH$_3$ & 233601.55 & 122 & 3.5806E-04 & 370 & $18_{(7,12)}-17_{(6,11)}$; AA \\
CH$_3$COCH$_3$ & 218127.21 & 119 & 2.2105E-04 & 656 & $20_{(3,18)}-19_{(2,17)}$; EE \\
CH$_3$COCH$_3$ & 218127.21 & 119 & 2.2105E-04 & 656 & $20_{(2,18)}-19_{(3,17)}$; EE \\
CH$_3$COCH$_3$ & 230182.81 & 220 & 4.2074E-05 & 912 & $28_{(3,26)}-28_{(2,27)}$; EE \\
CH$_3$COCH$_3$ & 230182.81 & 220 & 4.2074E-05 & 912 & $28_{(2,26)}-28_{(1,27)}$; EE \\
CH$_3$COCH$_3$ & 217022.51 & 115 & 3.8943E-04 & 624 & $19_{(3,16)}-18_{(4,15)}$; EE \\
CH$_3$COCH$_3$ & 217022.51 & 115 & 3.8943E-04 & 624 & $19_{(4,16)}-18_{(3,15)}$; EE \\
CH$_3$COCH$_3$ & 232962.37 & 115 & 2.9451E-04 & 210 & $17_{(8,10)}-16_{(7,9)}$; AA \\
CH$_3$COCH$_3$ & 218127.21 & 119 & 2.2105E-04 & 656 & $20_{(3,18)}-19_{(2,17)}$; EE \\
CH$_3$COCH$_3$ & 218127.21 & 119 & 2.2105E-04 & 656 & $20_{(2,18)}-19_{(3,17)}$; EE \\
CH$_3$COCH$_3$ & 230990.98 & 172 & 2.8660E-05 & 410 & $20_{(13,8)}-19_{(14,5)}$; AA \\
CH$_3$COCH$_3$ & 219242.14 & 122 & 3.8802E-05 & 688 & $21_{(2,20)}-20_{(1,19)}$; EE \\
CH$_3$COCH$_3$ & 219242.14 & 122 & 3.8802E-05 & 688 & $21_{(1,20)}-20_{(2,19)}$; EE \\
CH$_3$COCH$_3$ & 232181.92 & 381 & 1.0944E-05 & 976 & $30_{(19,11)}-30_{(16,14)}$; EE \\
CH$_3$COCH$_3$ & 218221.72 & 717 & 2.3951E-05 & 154 & $38_{(38,0)}-38_{(37,1)}$; AE \\
CH$_3$COCH$_3$ & 218221.72 & 717 & 2.3950E-05 & 462 & $38_{(38,1)}-38_{(37,2)}$; AE \\
CH$_3$COCH$_3$ & 219242.14 & 122 & 4.3148E-04 & 688 & $21_{(1,20)}-20_{(1,19)}$; EE \\
CH$_3$COCH$_3$ & 219242.14 & 122 & 4.3148E-04 & 688 & $21_{(2,20)}-20_{(2,19)}$; EE \\
CH$_3$COCH$_3$ & 218281.37 & 899 & 2.4303E-04 & 582 & $48_{(22,26)}-48_{(21,27)}$; AA \\
CH$_3$COCH$_3$ & 218281.38 & 899 & 2.4301E-04 & 970 & $48_{(23,26)}-48_{(22,27)}$; AA \\
CH$_3$COCH$_3$ & 217240.80 & 900 & 1.1949E-04 & 356 & $44_{(39,6)}-44_{(38,7)}$; EA \\
CH$_3$COCH$_3$ & 231738.68 & 922 & 9.7698E-05 & 356 & $44_{(41,4)}-44_{(40,5)}$; EA \\
CH$_3$COCH$_3$ & 215892.02 & 932 & 1.3305E-04 & 910 & $45_{(39,6)}-45_{(38,7)}$; AA \\
CH$_3$COCH$_3$ & 231199.96 & 765 & 3.7119E-04 & 1456 & $45_{(18,27)}-45_{(17,28)}$; EE \\
CH$_3$COCH$_3$ & 231199.96 & 765 & 3.7119E-04 & 1456 & $45_{(19,27)}-45_{(18,28)}$; EE \\
CH$_3$COCH$_3$ & 230183.08 & 135 & 5.8067E-04 & 282 & $23_{(1,23)}-22_{(0,22)}$; AA \\
CH$_3$COCH$_3$ & 215892.02 & 932 & 1.3303E-04 & 546 & $45_{(39,7)}-45_{(38,8)}$; AA \\
CH$_3$COCH$_3$ & 230183.08 & 135 & 5.8061E-04 & 470 & $23_{(0,23)}-22_{(1,22)}$; AA \\
CH$_3$COCH$_3$ & 230176.73 & 135 & 1.2709E-04 & 752 & $23_{(1,23)}-22_{(1,22)}$; EE \\
CH$_3$COCH$_3$ & 230176.73 & 135 & 1.2709E-04 & 752 & $23_{(0,23)}-22_{(0,22)}$; EE \\
CH$_3$COCH$_3$ & 233600.13 & 122 & 3.5802E-04 & 222 & $18_{(6,12)}-17_{(7,11)}$; AA \\
CH$_3$COCH$_3$ & 230176.73 & 135 & 1.2709E-04 & 752 & $23_{(1,23)}-22_{(1,22)}$; EE \\
CH$_3$COCH$_3$ & 230176.73 & 135 & 1.2709E-04 & 752 & $23_{(0,23)}-22_{(0,22)}$; EE \\
CH$_3$COCH$_3$ & 230176.73 & 135 & 4.5352E-04 & 752 & $23_{(1,23)}-22_{(0,22)}$; EE \\
CH$_3$COCH$_3$ & 230176.73 & 135 & 4.5352E-04 & 752 & $23_{(0,23)}-22_{(1,22)}$; EE \\
CH$_3$COCH$_3$ & 230182.81 & 220 & 4.2074E-05 & 912 & $28_{(3,26)}-28_{(2,27)}$; EE \\
CH$_3$COCH$_3$ & 230182.81 & 220 & 4.2074E-05 & 912 & $28_{(2,26)}-28_{(1,27)}$; EE \\
HNCO & 219798.27 & 58 & 1.4693E-04 & 21 & $10_{(0,10)}-9_{(0,9)}$ \\
HNCO & 218981.01 & 101 & 1.4222E-04 & 21 & $10_{(1,10)}-9_{(1,9)}$ \\
HNCO & 219733.85 & 228 & 1.3458E-04 & 21 & $10_{(2,9)}-9_{(2,8)}$ \\
HNCO & 219737.19 & 228 & 1.3458E-04 & 21 & $10_{(2,8)}-9_{(2,7)}$ \\
HNCO & 219656.77 & 433 & 1.2012E-04 & 21 & $10_{(3,8)}-9_{(3,7)}$ \\
HNCO & 219656.77 & 433 & 1.2012E-04 & 21 & $10_{(3,7)}-9_{(3,6)}$ \\
HNCO & 231873.26 & 470 & 6.6812E-05 & 57 & $28_{(1,28)}-29_{(0,29)}$ \\
$^{13}$CH$_3$CN & 232194.91 & 142 & 1.0217E-03 & 54 & $13_{3}-12_{3}$; A2-A1 \\
$^{13}$CH$_3$CN & 232194.91 & 142 & 1.0217E-03 & 54 & $13_{3}-12_{3}$; A1-A2 \\
$^{13}$CH$_3$CN & 232077.20 & 336 & 8.4805E-04 & 54 & $13_{6}-12_{6}$; A1-A2 \\
$^{13}$CH$_3$CN & 232077.20 & 336 & 8.4805E-04 & 54 & $13_{6}-12_{6}$; A2-A1 \\
$^{13}$CH$_3$CN & 232194.91 & 142 & 1.0217E-03 & 54 & $13_{3}-12_{3}$; A2-A1 \\
$^{13}$CH$_3$CN & 232194.91 & 142 & 1.0217E-03 & 54 & $13_{3}-12_{3}$; A1-A2 \\
C$_2$H$_5$CN & 231269.83 & 1056 & 1.4160E-05 & 117 & $58_{(17,42)}-59_{(16,43)}$ \\
C$_2$H$_5$CN & 231269.83 & 1056 & 1.4160E-05 & 117 & $58_{(17,41)}-59_{(16,44)}$ \\
C$_2$H$_5$CN & 233069.31 & 205 & 9.9390E-04 & 53 & $26_{(7,20)}-25_{(7,19)}$ \\
C$_2$H$_5$CN & 233069.38 & 205 & 9.9390E-04 & 53 & $26_{(7,19)}-25_{(7,18)}$ \\
C$_2$H$_5$CN & 233205.09 & 191 & 1.0164E-03 & 53 & $26_{(6,21)}-25_{(6,20)}$ \\
C$_2$H$_5$CN & 233498.30 & 179 & 1.0375E-03 & 53 & $26_{(5,21)}-25_{(5,20)}$ \\
C$_2$H$_5$CN & 233654.02 & 169 & 1.0540E-03 & 53 & $26_{(4,23)}-25_{(4,22)}$ \\
C$_2$H$_5$CN & 231312.30 & 158 & 9.1831E-05 & 55 & $27_{(0,27)}-26_{(1,26)}$ \\
C$_2$H$_5$CN & 232998.74 & 222 & 9.6916E-04 & 53 & $26_{(8,19)}-25_{(8,18)}$ \\
C$_2$H$_5$CN & 232998.74 & 222 & 9.6916E-04 & 53 & $26_{(8,18)}-25_{(8,17)}$ \\
C$_2$H$_5$CN & 232967.57 & 241 & 9.4197E-04 & 53 & $26_{(9,17)}-25_{(9,16)}$ \\
C$_2$H$_5$CN & 232967.57 & 241 & 9.4197E-04 & 53 & $26_{(9,18)}-25_{(9,17)}$ \\
C$_2$H$_5$CN & 232967.57 & 241 & 9.4197E-04 & 53 & $26_{(9,17)}-25_{(9,16)}$ \\
C$_2$H$_5$CN & 232967.57 & 241 & 9.4197E-04 & 53 & $26_{(9,18)}-25_{(9,17)}$ \\
C$_2$H$_5$CN & 231269.83 & 1056 & 1.4160E-05 & 117 & $58_{(17,42)}-59_{(16,43)}$ \\
C$_2$H$_5$CN & 231269.83 & 1056 & 1.4160E-05 & 117 & $58_{(17,41)}-59_{(16,44)}$ \\
C$_2$H$_5$CN & 232286.30 & 738 & 5.6326E-05 & 115 & $57_{(4,53)}-57_{(3,54)}$ \\
C$_2$H$_5$CN & 232962.32 & 262 & 9.1185E-04 & 53 & $26_{(10,16)}-25_{(10,15)}$ \\
C$_2$H$_5$CN & 232962.32 & 262 & 9.1185E-04 & 53 & $26_{(10,17)}-25_{(10,16)}$ \\
C$_2$H$_5$CN & 233523.50 & 592 & 4.4001E-04 & 53 & $26_{(20,7)}-25_{(20,6)}$ \\
C$_2$H$_5$CN & 233523.50 & 592 & 4.4001E-04 & 53 & $26_{(20,6)}-25_{(20,5)}$ \\
C$_2$H$_5$CN & 233523.50 & 592 & 4.4001E-04 & 53 & $26_{(20,7)}-25_{(20,6)}$ \\
C$_2$H$_5$CN & 233523.50 & 592 & 4.4001E-04 & 53 & $26_{(20,6)}-25_{(20,5)}$ \\
C$_2$H$_5$CN & 233208.03 & 434 & 6.6695E-04 & 53 & $26_{(16,10)}-25_{(16,9)}$ \\
C$_2$H$_5$CN & 233208.03 & 434 & 6.6695E-04 & 53 & $26_{(16,11)}-25_{(16,10)}$ \\
HCCCN & 218324.72 & 131 & 8.2613E-04 & 49 & $24-23$ \\
NH$_2$CHO & 233488.89 & 258 & 4.3644E-04 & 23 & $11_{(8,3)}-10_{(8,2)}$ \\
NH$_2$CHO & 233488.89 & 258 & 4.3644E-04 & 23 & $11_{(8,4)}-10_{(8,3)}$ \\
NH$_2$CHO & 233505.52 & 365 & 1.6086E-04 & 23 & $11_{(10,1)}-10_{(10,0)}$ \\
NH$_2$CHO & 233505.52 & 365 & 1.6086E-04 & 23 & $11_{(10,2)}-10_{(10,1)}$ \\
NH$_2$CHO & 233492.68 & 308 & 3.0630E-04 & 23 & $11_{(9,2)}-10_{(9,1)}$ \\
NH$_2$CHO & 233492.68 & 308 & 3.0630E-04 & 23 & $11_{(9,3)}-10_{(9,2)}$ \\
NH$_2$CHO & 233896.58 & 94 & 8.6203E-04 & 23 & $11_{(3,9)}-10_{(3,8)}$ \\
NH$_2$CHO & 233498.07 & 213 & 5.5141E-04 & 23 & $11_{(7,4)}-10_{(7,3)}$ \\
NH$_2$CHO & 233498.07 & 213 & 5.5141E-04 & 23 & $11_{(7,5)}-10_{(7,4)}$ \\
NH$_2$CHO & 233527.80 & 174 & 6.5125E-04 & 23 & $11_{(6,6)}-10_{(6,5)}$ \\
NH$_2$CHO & 233527.80 & 174 & 6.5125E-04 & 23 & $11_{(6,5)}-10_{(6,4)}$ \\
NH$_2$CHO & 233594.50 & 142 & 7.3618E-04 & 23 & $11_{(5,7)}-10_{(5,6)}$ \\
NH$_2$CHO & 233594.50 & 142 & 7.3618E-04 & 23 & $11_{(5,6)}-10_{(5,5)}$ \\
NH$_2$CHO & 233734.72 & 115 & 8.0662E-04 & 23 & $11_{(4,8)}-10_{(4,7)}$ \\
NH$_2$CHO & 232273.65 & 79 & 8.8167E-04 & 23 & $11_{(2,10)}-10_{(2,9)}$ \\
NH$_2$CHO & 215687.01 & 68 & 6.9810E-04 & 21 & $10_{(2,8)}-9_{(2,7)}$ \\
NH$_2$CHO & 218459.21 & 61 & 7.4751E-04 & 21 & $10_{(1,9)}-9_{(1,8)}$ \\
NH$_2$CHO & 233498.07 & 213 & 5.5141E-04 & 23 & $11_{(7,4)}-10_{(7,3)}$ \\
NH$_2$CHO & 233498.07 & 213 & 5.5141E-04 & 23 & $11_{(7,5)}-10_{(7,4)}$ \\
OCS & 218903.36 & 100 & 3.0384E-05 & 37 & $18-17$ \\
OCS & 231060.99 & 111 & 3.5783E-05 & 39 & $19-18$
\enddata
\end{deluxetable}

\section{Supplementary Images \label{appx:img}}
\resetapptablenumbers

\begin{figure*}[htb!]
\centering
\includegraphics[width=\linewidth]{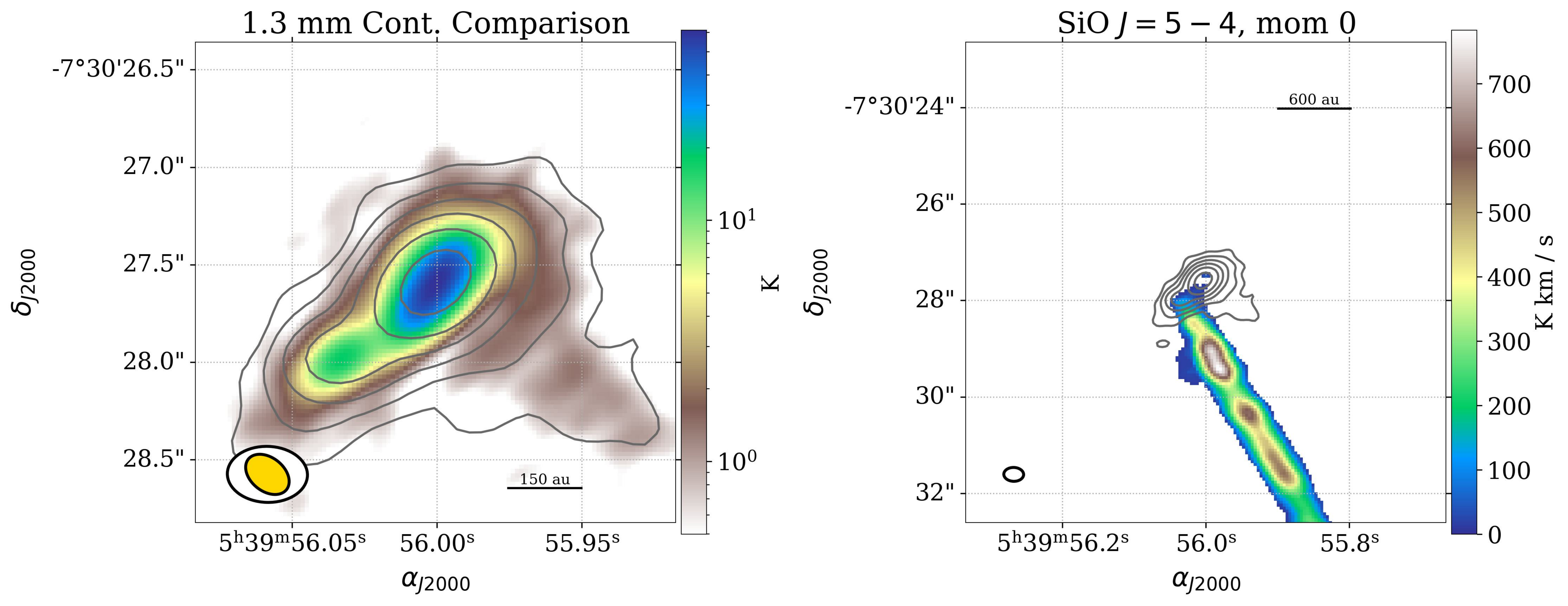}
\caption{\label{fig:appx_img} 
The selected images of dust continuum or integrated intensity of molecular transitions. 
The ellipses at the bottom left corner illustrate the observation beam(s), yellow for ALMA \#2018.1.01038.S (high resolution) and white for \#2018.1.00302.S (wide bandwidth).  
Panel (a): the comparisons of the 1.3~mm (230~GHz) dust continuum between the two programs used in this study. 
Comparisons of the 1.3~mm dust continuum observed by the high-resolution (color scale) and the wide-coverage (grey contours) data. 
The contour levels are [5, 10, 20, 40, 80, 160]$\sigma$, where $\sigma$ is 0.13 K. 
Pabel (b): The SiO $J=5-4$ integrated intensity (-15.5 -- 24.5 \kmpers) overlaid with 1.3~mm (230~GHz) dust continuum, both obtained from the wide coverage data. 
The contour levels are [5, 10, 20, 40, 80, 160]$\sigma$, where $\sigma$ is 0.13 K. 
A mask of 5$\sigma$ was applied during the integration. 
}
\end{figure*}

Figure~\ref{fig:appx_img} shows supplementary images of this study. 

\bibliography{REFERENCE.bib}{}
\bibliographystyle{aasjournal}




\end{CJK*}
\end{document}